\documentclass[12pt, fleqn]{article}
\usepackage[cp1251]{inputenc}
\usepackage{latexsym,amsfonts,amssymb}
\usepackage{graphicx}

\usepackage{amsbsy}
\usepackage{amsmath}
\usepackage{epsf}

\newtheorem{theo} {Theorem}
\newcommand{\bteo}{\begin{theo}}  
\newcommand{\et}{\end{theo}}
\newtheorem{Remark}{Remark}
\newtheorem{definition}{Definition}
\newcommand{\bd}{\begin{displaymath}}
\newcommand{\ed}{\end{displaymath}}

\newcommand{\lf}{\left}
\newcommand{\rg}{\right}
\newcommand{\be} {\begin{equation}}
\newcommand{\ee} {\end{equation}}
\newcommand{\ba} {\begin{array}{l}}
\newcommand{\ea} {\end{array}}
\newcommand{\bea}{\begin{eqnarray}}
\newcommand{\eea} {\end{eqnarray}}
\newcommand{\p} {\partial}

\begin{document}

\begin{center}
 {\Large \bf The  Shigesada--Kawasaki--Teramoto  model:  \\
conditional symmetries, exact solutions \\and their properties}

\medskip

{\bf Roman Cherniha $^{a,b,}$\footnote{\small  Corresponding author.
E-mail: r.m.cherniha@gmail.com; roman.cherniha1@nottingham.ac.uk},
Vasyl' Davydovych$^{a}$ and John R. King$^{b}$ }

$^{a}$ \quad Institute of Mathematics,  National Academy of Sciences
of Ukraine, \\
 3, Tereshchenkivs'ka Street, Kyiv 01004, Ukraine.\\
 $^{b}$ \quad School of Mathematical Sciences, University of Nottingham,\\
  University Park, Nottingham NG7 2RD, UK
\end{center}

\begin{abstract} We study a simplification of the well-known Shigesada--Kawasaki--Teramoto  model, which consists
 of two nonlinear reaction-diffusion equations with cross-diffusion.
 A complete  set of  $Q$-conditional (nonclassical) symmetries is derived using an algorithm adopted for the construction of
  conditional symmetries. The symmetries obtained are applied for finding  a wide range of exact solutions, possible biological interpretation of some
  of which being  presented. Moreover, an alternative application of the simplified model  related to the polymerisation process   is suggested
 and    exact solutions are found in this case  as well.
\end{abstract}

\section{Introduction} \label{sec-1}
Here we study      a  simplification of the well-known
Shigesada--Kawasaki--Teramoto (SKT) system  \cite{sh-ka-te}, which
is a nontrivial generalization of     the diffusive Lotka--Volterra
(DLV) system; we refer to \cite{sh-ka-1997} for general background.
The simplified SKT system   reads \be\label{2-1}
 \ba
  u_t = [(d_1+d_{12}v)u]_{xx}+u(a_1-b_1u-c_1v),\\
 v_t = [(d_2+d_{21}u)v]_{xx}+v(a_2-b_2u-c_2v),
 \ea
  \ee
where   the lower subscripts  $t$ and $x$ mean differentiation with
respect to these variables,  $u=u(t,x)$  and $v=v(t,x)$ are two
unknown functions, which usually  represent densities of two
competing species, $d_1$ and  $d_2$ denote the standard  diffusion
coefficients, $d_{12}v$ and $d_{21}u$ are cross-diffusion pressures,
$a_1$ and $a_2$ are  the intrinsic growth coefficients, $b_1$ and
$c_2$ denote the coefficients of intra-specific competitions and
$b_2$ and $c_1$ denote the coefficients of inter-specific
competitions. Notably, the  simplified SKT system (\ref{2-1}) was
under study in \cite{lou-2017} (with $d_{21}=0$) and
\cite{kan-2020}.

 Hereafter $a_k$, $b_k,$ $c_k$, $d_k$ ($k=1,2$),  $d_{12}$ and $d_{21}$ are assumed to be real constants and the restrictions
\be\label{2-2}
  d^2_{12}+d^2_{21}\not=0, \ d_1^2+d_{12}^2\not=0, \ d_2^2+d_{21}^2\not=0, \ d_1^2+d_{2}^2\not=0
   \ee and \be\label{2-2*}
  d_{12}^2+c_1^2\not=0, \ d_{21}^2+b_2^2\not=0
   \ee hold.
   The first restriction from (\ref{2-2}) means that we look for  SKT systems that  are not  reducible  to those without cross diffusion.
    The remaining three conditions in  (\ref{2-2})  guarantee  that both equations of system (\ref{2-1}) are  second-order
    partial differential equations (PDEs).
    In particular, if one sets  $d_1=d_{2}=0$ then the system obtained  is reducible
   via  the transformation \[ u \rightarrow \frac{u}{d_{21}}, \ v \rightarrow \frac{v}{d_{12}}\] to the  system \be\nonumber
 \ba
  u_t-v_t = u(a_1-b_1^*u-c_1^*v)-v(a_2-b_2^*u-c_2^*v),\\
 v_t = [uv]_{xx}+v(a_2-b_2^*u-c_2^*v),
 \ea
  \ee
  involving   an ordinary differential equation (ODE)
  (here parameters with stars are easily  determined).
 Finally, two restrictions  (\ref{2-2*})  are adopted   in order to   exclude
    systems of the form (\ref{2-1}) that are not coupled, so that
   interactions between species (cells, chemicals) are captured.

 The nonlinear system (\ref{2-1}) is a special case  of the SKT system, which was
initially  proposed in the form  \cite{sh-ka-te}
 \be\label{2-1*}
 \ba
  u_t = [(d_1+d_{11}u+d_{12}v)u]_{xx}+ (W_xu)_x+u(a_1-b_1u-c_1v),\\
 v_t = [(d_2+d_{21}u+d_{22}v)v]_{xx}+(W_xv)_x+v(a_2-b_2u-c_2v),
 \ea
  \ee
  where the function $W(x)$ is the so-called environmental potential, which typically is assumed to be a constant or a linear function (in both cases the convective terms are removable).
Nowadays,  many works are devoted to investigation of the
 SKT system (\ref{2-1*}) and its particular cases  by
different qualitative and analytical  methods, see, e.g., the recent
papers
\cite{lou-2017,kan-2020,pham-2017,pham-2019,wu-2020,kersner-2022}
and references cited therein.
Several studies are devoted to solving this system using different
numerical  techniques  to construct  numerical solutions
\cite{andreianov-2011,berres-2011,gambino-2012,li-2020}.

 However, we concentrate ourselves here
on papers devoted to the search for exact solutions of the SKT
system (\ref{2-1*}). The majority of such papers are devoted to
symmetry-based methods, which can be described as follows:
 in order to find  exact  solutions of a nonlinear PDE (system of PDEs) one solves the equation (system) in question
       together with a differential
constraint(s) generated by a symmetry operator. The corresponding
symmetry can be of different types. The most common   for solving
nonlinear PDEs are  Lie symmetry (there is an enormous number of
excellent papers devoted to classical  symmetries, hence only recent
monographs are cited here)
\cite{bl-anco-10,arrigo-15,ch-se-pl-book}, $Q$-conditional
(nonclassical) symmetry \cite{bl-anco-10,bl-co-69, ch-dav-book,
oliveri-21}, conditional  symmetry \cite{Fush88,fss, ch-hen-2010},
generalized conditional symmetry \cite{Foka94, qu-97} (the
terminology `conditional Lie--B\"acklund symmetry'
\cite{zh95,ji-qu-2013} is sometimes used for the same symmetry
type).

Lie symmetries of the SKT system   are completely described in
\cite{ch-dav-muzy-17}, and examples of  exact solutions are
presented therein as well. In  \cite{ch-myr-08}, the method of
additional generating conditions is applied to obtain new non-Lie
ans\"atze reducing the SKT system (\ref{2-1*}) to ODE systems of and
to find new exact solutions.
 Recently, a wide range of
generalized conditional symmetries of this system were
 identified  in \cite{li-2020,li-2022};
  however, the authors were unable to find nontrivial exact solutions using the symmetries obtained.
  To the best of our knowledge, there are no other studies devoted to the search for symmetries and/or  exact solutions of  the SKT system (\ref{2-1*}).

 In particular, $Q$-conditional (nonclassical) symmetry of the SKT system   was still unknown.
 Thus, the main aim of this study is construction of this type of symmetries and their application for finding
 exact solutions. It should be mentioned that
 the algorithm for  construction of  $Q$-conditional  symmetries is similar to that for Lie symmetries.
 However, the latter  leads to a so-called  system of determining equations, which   is always linear, while
 one needs to solve a nonlinear system of determining equations in order to find $Q$-conditional  symmetries. Thus,  it is always  a  challenge because
  the relevant  nonlinear system can be nonintegrable. Here it is shown that the system of determining equations for the SKT system (\ref{2-1})
 is integrable, so that a complete list of possible  $Q$-conditional  symmetries is derived.
 Moreover, we demonstrate that the $Q$-conditional  symmetries obtained can be successfully
 used for finding new exact solutions of system (\ref{2-1}).

 It should be mentioned that the SKT system (\ref{2-1*}) with $d_{ij}=0 \ (i,j=1,2) $ coincides with the DLV system. The
  Lie symmetries
  of the latter are presented in \cite{ch-du-04} (they can be also identified from the  more general results derived in
  \cite{ch-king1,ch-king2}). It was shown in \cite{ch-dav-muzy-17} that the Lie symmetry of  the SKT system (\ref{2-1*})
    differs essentially from that of the DLV system;
   $Q$-conditional (nonclassical) symmetries of  the DLV system
   were constructed  as well and extensively used for constructions of exact solutions
    (see  the  recent review  \cite{ch-dav-2022} and references cited therein).
    We show here that $Q$-conditional  symmetries of  (\ref{2-1}) are  not obtainable
     from those of the DLV system.
%

The scalar analogue of  the   SKT system (\ref{2-1}) can be written
as
 \[ u_t = [(d_1+d_{12}u)u]_{xx}+u(a_1-b_1u),
  \]
  furnishing solutions to (\ref{2-1}), in which $u$ is proportional to $v$ for special choices of the parameters.
  In contrast to the vector case, the above equation  can be easily rewritten in the  divergence  form
  \be\label{1-1}
  u_t = [(d_1+2d_{12}u)u_x]_{x}+u(a_1-b_1u).
  \ee
  Now one notes that Eq.\,(\ref{1-1}) with  $d_{12}=0$ is the famous Fisher equation,
  while  that with $d_1=0$ is referred to as  the porous-Fisher equation  \cite{mur-89,witel-95, fadai-2020}. Lie and $Q$-conditional symmetries of the Fisher  equation  are  trivial provided $a_1$ and $b_1$ are
   arbitrary nonzero parameters (see, e.g.,  \cite{ch-se-pl-book}).
   The porous-Fisher possesses a nontrivial Lie symmetry (see case 24 in Table 2.5 \cite{ch-se-pl-book}).
   A nontrivial $Q$-conditional  symmetry  was also found \cite{ar-hill-95};
    however, this is equivalent to the corresponding Lie symmetry. Finally, assuming $d_1d_{12}\not=0$, it can
     be checked that Eq.\,(\ref{1-1})
           admits a nontrivial symmetry only in
            the case in which it is reducible to the porous-Fisher
            equation. Although  Eq.\,(\ref{1-1}) with $d_1d_{12}\not=0$ possesses
             `poor' symmetry, highly nontrivial exact solutions have been  found, for example by  using
             the method of additional generating conditions \cite{ch-2001}.

The paper is organized as follows. In Section~\ref{sec-2}, a
complete description of the $Q$-conditional symmetries
 of the simplified SKT system (\ref{2-1}) is provided
 (the  so-called `no go' case is not examined here).
 In Section~\ref{sec-3}, we present some examples of exact solutions and their possible interpretation.
 In Section~\ref{sec-4}, an alternative application of  system (\ref{2-1}) is presented and further exact solutions are constructed.  Finally, we  discuss the results obtained and  present some conclusions  in
the last section. In particular, a comparison of the results
obtained  with those derived in other papers is presented.

\section{Main results} \label{sec-2}

It can be noted that the SKT system (\ref{2-1*}) admits a highly
nontrivial Lie symmetry only in the cases when at least one
cross-diffusion coefficient $d_{ij}, \ i\not=j$ vanishes
\cite{ch-dav-muzy-17} (see Cases 5--7, 9--16 of Table 1 therein).
This means that  essential features of the model are lost in these
cases. Cases~5 and 8 of Table 1 \cite{ch-dav-muzy-17} involve both
cross-diffusivities; however, the reaction terms are linear, i.e the
model   does not predict interaction between species. Of course,
Cases 1--3 of Table 1 \cite{ch-dav-muzy-17} present specific  forms
of the SKT system, which do not affect the biological sense of this
model, and the relevant systems admit nontrivial Lie symmetries.
Here we demonstrate that the $Q$-conditional (nonclassical) symmetry
of  the SKT system (\ref{2-1*}) is much wider, hence the symmetries
obtained  essentially extend possibilities for constructing new
exact solutions in the cases  that may be  important from the
applicability point of view.

Let us  consider the general form of $Q$-conditional symmetry
operator of system (\ref{2-1})\,: \be \label{2-33}
 Q=\xi^0(t,x,u,v)\partial_t+\xi^1(t,x,u,v)\partial_x+\eta^1(t,x,u,v)\partial_u+\eta^2(t,x,u,v)\partial_v, \ \lf(\xi^0\rg)^2+\lf(\xi^1\rg)^2\neq0,
\ee where    $\xi^0, \ \xi^1, \ \eta^1$ and $\eta^2$ are
to-be-determined smooth functions.

First of all, we note that the SKT system (\ref{2-1}) is a system of
evolution equations. Therefore, the problem of constructing its
$Q$-conditional symmetry operators of the form (\ref{2-33})
essentially depends on the value of the function $\xi^0$. Thus, one
needs to consider two different cases\,:
\begin{enumerate}
  \item $\xi^0\neq0;$
  \item $\xi^0=0, \ \xi^1\neq0.$
\end{enumerate}

It is well-known that the problem of finding $Q$-conditional
symmetry operators with $\xi^0=0$ (i.e. the `no go' case)  for a
single evolution PDE  is equivalent to that of the construction of
the general solution for the given PDE (see   \cite{zh-lah-98} for
details). Thus, this problem cannot be completely solved for any
nonintegrable evolution equation. For this reason, we restrict
ourselves only to the consideration of Case 1. Thus, in this case
operator (\ref{2-33}) can be rewritten as
  \begin{equation} \label{2-29}\begin{array}{l}
 Q=\partial_t+\xi(t,x,u,v)\partial_x+\eta^1(t,x,u,v)\partial_u+\eta^2(t,x,u,v)\partial_v.
\end{array}\end{equation}

\begin{definition} \label{d2}
Operator (\ref{2-29}) is called the $Q$-conditional symmetry for
system (\ref{2-1}) if  the following invariant conditions are
satisfied\,:
\[\ba
\mbox{\raisebox{-1.6ex}{$\stackrel{\displaystyle  
Q}{\scriptstyle 2}$}}  
\lf(S_1\rg)  
\Big\vert_{{\cal{M}}}=0, \
\mbox{\raisebox{-1.6ex}{$\stackrel{\displaystyle  
Q}{\scriptstyle 2}$}}  
\lf(S_2\rg)  
\Big\vert_{{\cal{M}}}=0, \ea
\]
 where $\mbox{\raisebox{-1.7ex}{$\stackrel{\displaystyle Q}{\scriptstyle
2}$}}$ is the second prolongation of the operator $Q$,  the manifold
 \[{\cal{M}}=\{S_1=0,S_2=0,Q(u)=0,Q(v)=0\},\] with
 \be\nonumber\ba
 S_1 \equiv \  [(d_1+d_{12}v)u]_{xx}-u_t+u(a_1-b_1u-c_1v),\\
S_2 \equiv \   [(d_2+d_{21}u)v]_{xx}-v_t+v(a_2-b_2u-c_2v).\ea\ee
\end{definition}

Hereafter we use the notations \[\ba Q(u)=u_t+\xi u_x-\eta^1, \ Q(v)=v_t+\xi v_x-\eta^2, \medskip \\
\mbox{\raisebox{-1.7ex}{$\stackrel{\displaystyle Q}{\scriptstyle
2}$}}=Q+\rho^1_t\p_{u_{t}}+\rho^1_x\p_{u_{x}}+\rho^2_t\p_{v_{t}}+\rho^2_x\p_{v_{x}}+\medskip\\
\hskip2cm \sigma^1_{tt}\p_{u_{tt}}
+\sigma^1_{tx}\p_{u_{tx}}+\sigma^1_{xx}\p_{u_{xx}}+\sigma^2_{tt}\p_{v_{tt}}
+\sigma^2_{tx}\p_{v_{tx}}+\sigma^2_{xx}\p_{v_{xx}},\ea\] where the
coefficients $\rho$ and $\sigma$ with relevant subscripts are
expressed via the functions $\xi^i$ and $\eta^k$ by  well-known
formulae (see, e.g., \cite{bl-anco-10,fss}).

\begin{Remark}
Formally speaking, the manifold ${\cal{M}}$ should contain also the
first-order differential consequences of the equations $Q(u)=0$ and
$Q(v)=0$
 according to the general definition of conditional symmetry (see Chapter 5 \cite{bl-anco-10}).
 However, we have shown that these consequences can be skipped in the case of
   evolution systems (see Chapter 2 \cite{ch-dav-book}). Obviously, (\ref{2-1}) belongs to    systems of that type.
\end{Remark}

Applying Definition~\ref{d2} and making straightforward
calculations, we establish that the most general form of
$Q$-conditional symmetry operator (\ref{2-29})  for system
(\ref{2-1}) is \be\label{2-35}
Q=\p_t+\xi(t,x)\,\p_x+\eta^1(t,x,u,v)\,\partial_u+\eta^2(t,x,u,v)\,\partial_v,\ee
where the functions $\xi, \ \eta^1$ and $\eta^2$  should be found
from the \emph{system of determining equations} \begin{small}
\begin{eqnarray} \label{2-36}
 &&  D\eta^1_{vv}+2d_2d_{12}\eta^1_v=0, \ D\eta^2_{uu}+2d_1d_{21}\eta^2_u=0, \\
 \label{2-37} && D\eta^1_{uu}+2d_2d_{12}\eta^2_u=0, \
  D^2\eta^1_{uv}=d_2d_{12}\left(d_1d_{21}\eta^1+d_2d_{12}\eta^2\right)+D\left(d_1d_{21}\eta^1_v-d_2d_{12}\eta^2_v\right),  \\
  \label{2-38}  && D\eta^2_{vv}+2d_1d_{21}\eta^1_v=0, \ D^2\eta^2_{uv}=d_1d_{21}\left(d_1d_{21}\eta^1+
  d_2d_{12}\eta^2\right)-D\left(d_1d_{21}\eta^1_u-d_2d_{12}\eta^2_u\right), \\
   \nonumber &&\\
    &&\hskip1cm 2D^2\eta^1_{xu}=(d_2+d_{21}u)\left(d_2d_{12}\xi\eta^2-D\xi_t-2D\xi\xi_x\right)-
  d_2d_{12}d_{21}v\xi\eta^1+D^2\xi_{xx}-\nonumber\\ \label{2-39} && \hskip3cm d_{21}vD\xi\eta^1_{v}+
  d_{12}uD\xi\eta^2_{u}-2d_2d_{12}D\eta^2_{x}, \\
   \nonumber && \hskip1cm
   2D\eta^1_{xv}=\frac{d_2d_{12}(d_1+d_{12}v)}{D}\xi\eta^1-\frac{d_2d_{12}^2}{D}u\xi\eta^2-2d_2d_{12}\eta^1_x+
   \\ \label{2-41} && \hskip3cm (d_1-d_2-d_{21}u+d_{12}v)\xi\eta^1_v+d_{12}u\left(\xi_t+2\xi\xi_x-\xi\eta^1_u+\xi\eta^2_v\right) ,\\
  && \hskip1cm 2D^2\eta^2_{xv}=(d_1+d_{12}v)\left(d_1d_{21}\xi\eta^1-D\xi_t-2D\xi\xi_x\right)-
  d_1d_{12}d_{21}u\xi\eta^2+D^2\xi_{xx}+\nonumber\\
   \label{2-40} && \hskip3cm d_{21}vD\xi\eta^1_{v}-d_{12}uD\xi\eta^2_{u}-2d_1d_{21}D\eta^1_{x}, \\
    \nonumber && \hskip1cm
   2D\eta^2_{xu}=-\frac{d_1d_{21}^2}{D}v\xi\eta^1+\frac{d_1d_{21}(d_2+d_{21}u)}{D}\xi\eta^2-2d_1d_{21}\eta^2_x+
   \\ \label{2-42} && \hskip3cm (d_2-d_1+d_{21}u-d_{12}v)\xi\eta^2_u+d_{21}v\left(\xi_t+2\xi\xi_x+\xi\eta^1_u-\xi\eta^2_v\right) ,\\
   \nonumber &&\\
     \nonumber &&\hskip3cm 2D\left(\eta^1-F^1\right)\xi_x+ D\left(\eta^1_t-(d_1+d_{12}v)\eta^1_{xx}-d_{12}u\eta^2_{xx}\right)+\\
     \nonumber &&
     \Big[d_1d_{21}u\eta^1+(d_1+d_{12}v)(d_2F^1-u\Delta)\Big]\eta^1_u+
      \Big[D\eta^2+(d_1+d_{12}v)(d_1(F^2-\eta^2)+v\Delta)\Big]\eta^1_v+ \\
     \nonumber && \hskip3cm d_{12}u\Big[d_2(F^1-\eta^1)-u\Delta\Big]\eta^2_u+
     d_{12}u\Big[d_1(F^2-\eta^2)+v\Delta\Big]\eta^2_v+\\
     \label{2-43} &&
    d_{12}(v\eta^1-u\eta^2)\Delta- d_{12}(d_1+d_2)\eta^1\eta^2+\Big[d_1d_{12}F^2-DF^1_u\Big]\eta^1+\Big[c_1uD+d_2d_{12}F^1\Big]\eta^2=0,\\
      \nonumber && \hskip3cm 2D\left(\eta^2-F^2\right)\xi_x+ D\left(\eta^2_t-d_{21}v\eta^1_{xx}-(d_2+d_{21}u)\eta^2_{xx}\right)+\\
     \nonumber &&\hskip3cm d_{21}v\Big[d_2(F^1-\eta^1)-u\Delta\Big]\eta^1_u+d_{21}v\Big[d_1(F^2-\eta^2)+v\Delta\Big]\eta^1_v+
    \\
     \nonumber &&  \Big[D\eta^1+(d_2+d_{21}u)(d_2(F^1-\eta^1)-u\Delta)\Big]\eta^2_u+
     \Big[d_2d_{12}v\eta^2+(d_2+d_{21}u)(d_1F^2+v\Delta)\Big]\eta^2_v+
      \\
     \label{2-44} &&
     d_{21}(v\eta^1-u\eta^2)\Delta-d_{21}(d_1+d_2)\eta^1\eta^2+\Big[b_2vD+d_1d_{21}F^2\Big]\eta^1+\Big[d_2d_{21}F^1-DF^2_v\Big]\eta^2=0.
\end{eqnarray} \end{small}  where $D=d_1d_2+d_1d_{21}u+d_2d_{12}v, \ F^1=u(a_1-b_1u-c_1v), \
F^2=v(a_2-b_2u-c_2v),$  and $\Delta
=d_{12}(F^2-\eta^2)-d_{21}(F^1-\eta^1)$.

Completely solving the above system of determining equations leads
to the following theorem.

\begin{theo} \label{th3}
A system of the form (\ref{2-1}) with the  restrictions (\ref{2-2})
and (\ref{2-2*}) is invariant under  $Q$-conditional symmetry
ope\-ra\-tor(s) (\ref{2-29})
 if and only if this system  and the
corresponding operator(s)  have the forms listed in
  Table~\ref{tab1}. Any other system of the form (\ref{2-1}) admitting
  $Q$-conditional symmetry is reduced to the relevant system listed in
   Table~\ref{tab1}
  by  the  equivalence  transformations
\begin{equation}\label{2-31} t^*=\gamma_1t, \ x^*=\gamma_2x,  \ u^*=\gamma_3 u+\gamma_4 v, \
  v^*=\gamma_5 v+\gamma_6 u,\ee with correctly-specified constants  $\gamma_i$, $i=1,\dots,6$. \end{theo}

\emph{The proof} of the theorem is presented in Appendix \ref{A}.

\begin{small}
 \begin{table}
\caption{$Q$-conditional symmetries ofthe SKT  system
(\ref{2-1})}\medskip
\label{tab1}       
\begin{tabular}{|c|c|c|}
\hline  & The SKT systems &   Conditional symmetries and
restrictions   \\  \hline

1 & $u_t = [(1+d_{12}v)u]_{xx}+u\left(a_1-c_1v\right)$
 & $Q^1=\p_t+v(a_2-v)\,\p_v+
      $
 \\ &$v_t = [uv]_{xx}+v(a_2-b_2u-v)$&$u\Big(v+a_1+b_2(a_2d_{12}+1)-a_2(c_1+1)\Big)\,\p_u$ \\ \hline

2 & $u_t = [(1+v)u]_{xx}+u(a_1-b_1u-b_2v)$
 &  $b_1\neq0,
\  Q^2=\p_t+a_2v\,\p_v,$
 \\ &$ v_t = [uv]_{xx}+v(a_2-b_2u)$&$Q_0^3=\p_t+\frac{e^{a_2t}\,f(x)}{u}\,\p_v$
\\ \hline

3 & $u_t = [(1+d_{12}v)u]_{xx}+u(a_1-c_1v)$
 &   $Q^4=\p_t+\alpha\, u\,\p_u$
 \\ &$v_t = [uv]_{xx}-b_2uv$&
\\  \hline

4 &
$u_t = [(d_1+v)u]_{xx}-b_2uv$ &  $a_2+b_2d_2\neq0,$ \\
& $v_t = [(d_2+u)v]_{xx}+v(a_2-b_2u)$ &
$Q^5=\p_t+(a_2+b_2d_2)\big((d_2+u)\,\p_u+v\,\p_v\big)$\\  \hline

5 & $u_t =
[(d_1+v)u]_{xx}+u\left(-\frac{b_2d_1^2}{d_1+d_2}-b_2v\right)$
 &$b_2\neq0,  \ d_1+d_2\neq0, \ Q^6=\p_t+$\\
     & $v_t = [(d_2+u)v]_{xx}+v\left(-\frac{b_2d_2^2}{d_1+d_2}-b_2u\right)$&
    $\frac{b_2d_1d_2}{d_1+d_2}\left(\left(\frac{d_2^2}{d_1+d_2}+u\right)\p_u+
\left(\frac{d_1^2}{d_1+d_2}+ v\right) \p_v\right)$
 \\ \hline

6 & $u_t = [(1+v)u]_{xx}+u\left(a_1+a_1\,v\right)$
 &$ Q_0^7=\p_t+\beta u(1+v)\,\p_u+(a_2v-\beta v^2)\,\p_v,$
  \\ &  $v_t = [uv]_{xx}+v\left(a_2-b_2u-\left(a_1+b_2\right)v\right)$&
  $Q^7=\p_t+\beta u(1+v)\,\p_u+$ \\
  &&$(\beta+a_2+a_2v-\beta v^2)\,\p_v
  $ \\
 \hline

7 & $u_t = [(1+v)u]_{xx}+u\left(a_1-b_2\,v\right)$ & $a_1\neq -b_2,\
a_2\neq0,\ Q_0^{3},$
 \\ &$v_t = [uv]_{xx}+v(a_2-b_2u)$& $Q^{8}=\p_t+\alpha\,u\,\p_u+ a_2v\,\p_v$
 \\ \hline

8 & $u_t = [(1+v)u]_{xx}+u\left(a_1+(a_1+2a_2)\,v\right)$
&$a_2\neq0, \ Q^9=\p_t-2a_2\,u\,\p_u- a_2(1+v)\,\p_v, $
 \\ &$v_t = [uv]_{xx}+v(a_2+(a_1+2a_2)\,u+a_2\,v)$& $Q^{10}=\p_t-a_2u(3+v)\,\p_u+
  a_2(1+v)v\,\p_v$
 \\ \hline

9 & $u_t = [(1+v)u]_{xx}+u\left(a_1-b_2\,v\right)$ &$
 Q_0^{11}=\p_t+\alpha\,u\,\p_u+\frac{e^{2\alpha t}\,f(x)}{u}\,\p_v, \ Q^{11}=\p_t+$
 \\ &$v_t = [uv]_{xx}-b_2uv$&
 $\beta u\,\p_u+\frac{\beta e^{\beta t}}{\alpha_0e^{\beta t}+\beta-\alpha_0}
  \Big(\alpha_0(1+v)+\frac{ e^{\beta t}f(x)}{u}\Big)\,\p_v $
 \\ \hline

10 & $u_t = [(1+v)u]_{xx}-u\left(b_2+b_2\,v\right)$ &$a_2\neq0, \
Q^{8},$
 \\ &$v_t = [uv]_{xx}+v(a_2-b_2u)$& $Q^{3}=\p_t+\frac{a_2e^{a_2 t}}{\alpha_0e^{a_2
t}+a_2-\alpha_0}\left(\alpha_0(1+v)+
   \frac{f(x)}{u}\right)\,\p_v$
 \\ \hline

11 & $u_t = [(1+v)u]_{xx}-u\left(b_2+b_2\,v\right)$ &$Q_0^{11}, \
Q^{12}=\p_t+
   \left(\frac{1+v}{t}+\frac{f(x)}{tu}\right)\,\p_v$
 \\ &$v_t = [uv]_{xx}-b_2uv$&
 \\ \hline

12 & $u_t = [(1+v)u]_{xx}$
  & $Q^{13}=\p_t+\alpha\,\p_u+\alpha\,\p_v$
  \\ &$v_t = [(-1+u)v]_{xx}$& \\ \hline
 \end{tabular}
 \end{table}
 \end{small}

\begin{Remark} In Table~\ref{tab1},  $\alpha$ and $\alpha_0$ are  arbitrary constants,
$a_1+b_2=\beta\neq0, $
 while the
function $f(x)$ is the general solution of the linear ODE $f''-b_2f
=0$, i.e.
  \be\label{2-45}f(x)=\left\{
\begin{array}{l}
\alpha_1\exp(\sqrt{b_2}\,x)+\alpha_2\exp(-\sqrt{b_2}\,x), \
\mbox{if} \ b_2>0,  \medskip \\  \medskip
\alpha_1\cos(\sqrt{-b_2}\,x)+\alpha_2\sin(\sqrt{-b_2}\,x), \ \mbox{if} \ b_2<0,\\
 \alpha_1+\alpha_2x, \  \mbox{if} \  b_2=0.\\
  \end{array} \right. \ee
\end{Remark}

\begin{Remark}
The Lie symmetries of the  systems listed in Table~\ref{tab1} are
trivial (being formed by the operators of   time and space
translations) provided the
 coefficients are arbitrary and obey only  the restrictions given in Theorem~\ref{th3}.
 The only exception is the pure diffusion system in Case 12, which admits additionally
 the Lie symmetry operator of scale transformations (see system (45) in
  \cite{ch-dav-muzy-17}). This diffusion system implicitly requires $u>1$, otherwise the diffusivity is negative.
\end{Remark}

Using the equivalence transformations (\ref{2-31}), one can reduce
some parameters in the systems and the corresponding $Q$-conditional
symmetries
 presented in Table~\ref{tab1} to the values $\pm 1$.

For example, applying the transformation
\[ t^*=b_1\,t, \ x^*=\sqrt{b_1}\,x  \]
to the system arising in Case 2 of  Table~\ref{tab1} with $b_1>0$,
one obtains the system \be\label{2-1ad}
 \ba
u_{t^*} = [(1+v)u]_{x^*x^*}+u(\frac{a_1}{b_1}-u-\frac{b_2}{b_1}v), \\
v_{t^*} = [uv]_{x^*x^*}+v(\frac{a_2}{b_1}-\frac{b_2}{b_1}u). \ea\ee
 Thus, using the notations $a_1^*=\frac{a_1}{b_1}, \ a_2^*= \frac{a_2}{b_1}$ and $b_2^*= \frac{b_2}{b_1}$, one realizes that system  (\ref{2-1ad}) is nothing else but the relevant  system from Table~\ref{tab1} with $b_1=1$.
  All such cases have been identified and they are listed in Table~\ref{tab2}.
  Notably, some parameters are reducible to $\pm 1$ in Cases $1, 3, 4, 6, 9$ and $11$ as well, provided  the additional  assumption holds that a parameter  in question does not vanish. For example, the system arising in Case 11 is reducible to the system
 \be\label{2-2ad}
 \ba
u_{t^*} = [(1+v)u]_{x^*x^*}\mp u(1+v), \\
v_{t^*} = [uv]_{x^*x^*}\mp vu, \ea\ee by the transformation
\[ t^*=|b_2|\,t, \ x^*=\sqrt{|b_2|}\,x,  \]
under the assumption $b_2\not=0$. However, that system with $b_2=0$
is not reducible to the form (\ref{2-2ad}).


\begin{table}
\caption{Application of equivalence transformations }\medskip
\label{tab2}       
\begin{center}
\begin{tabular}{|c|c|c|}
\hline Case of Table~\ref{tab1} & Parameter &  Transformation  \\
\hline

2 & $b_1 \to \pm 1$ & $t^*=|b_1|\,t, \ x^*=\sqrt{|b_1|}\,x$
 \\ \hline

5 & $b_2 \to \pm 1$ & $t^*=|b_2|\,t, \ x^*=\sqrt{|b_2|}\,x$
 \\ \hline

7 & $a_2 \to \pm 1$ & $t^*=|a_2|\,t, \ x^*=\sqrt{|a_2|}\,x$
 \\ \hline

8 & $a_2 \to \pm 1$ & $t^*=|a_2|\,t, \ x^*=\sqrt{|a_2|}\,x$
 \\ \hline

10 & $a_2 \to \pm 1$ & $t^*=|a_2|\,t, \ x^*=\sqrt{|a_2|}\,x$
 \\ \hline
 \end{tabular}
 \end{center}
 \end{table}

From the symmetry point of view, the systems presented in
Table~\ref{tab1}
 can be divided into three categories. The first category consists of
  the systems admitting exactly one $Q$-conditional symmetry
   and they are presented in Cases 1, 4 and 5. The second one
   consists of the systems admitting a $Q$-conditional symmetry
   involving an arbitrary parameter, i.e. an one-parameter set of
   such symmetries, see Cases 3 and 12. Finally, Cases 2 (with the restriction $a_2\not= 0$),
   6--11 form the third category, in which the systems admit two $Q$-conditional symmetry
    with essentially different structures. Moreover, these symmetry operators
    depend on one or two arbitrary parameters (except that Cases 6 and 8 admit  two $Q$-conditional symmetries without arbitrary parameters).

The following observation concerning the conditional symmetry
operators  in Case 2 should be highlighted.
 Calculating Lie brackets of the
operators $Q^2$ and $Q_0^3$,   one obtains zero. However, the set
$\{Q^2, \ Q_0^3 \}$ does not form a Lie algebra because linear
combinations of these operators do not  produce $Q$-conditional
symmetry operators.
 Notably, Lie brackets of
the conditional symmetry operators  in all other cases (Cases 6--11)
produce  new operators, which are not conditional symmetries of the
relevant SKT system. We point out that this observation is in
agreement with the $Q$-conditional symmetry theory stating that,
generally  speaking,  the set of such symmetries for a given PDE
(system of PDEs) does not produce a Lie algebra
\cite{bl-anco-10,ch-se-pl-book}.

In conclusion of this section, let us  consider the SKT system
(\ref{2-1}) with $b_1=c_2=0$, i.e. \be\label{2-46}
 \ba
  u_t = [(d_1+d_{12}v)u]_{xx}+u(a_1-c_1v),\\
 v_t = [(d_2+d_{21}u)v]_{xx}+v(a_2-b_2u).
 \ea
  \ee
It is important to stress that many special cases of system
(\ref{2-46}) are listed in Table \ref{tab1} (see Cases 3--5, 7,
9--12). It turns out that the above system is doubly  exceptional
because it can be written in the conserved forms
\[
 \ba  \frac{\p}{\p t}\Big(\Phi u\Big)=  \frac{\p}{\p x}\Big(\Phi((d_1+d_{12}v)u)_{x}   - \Phi_x(d_1+d_{12}v)u\Big),\\
  \frac{\p}{\p t}\Big(\Psi v\Big)=   \frac{\p}{\p x}\Big(\Psi((d_2+d_{21}u)v)_{x}   - \Psi_x(d_2+d_{21}u)v\Big),
  \ea\]
  provided the functions
 \[ \Phi(t,x)= \exp\Big(\pm \sqrt{\frac{c_1}{d_{12}}}x-\Big(a_1+\frac{c_1d_1}{d_{12}}\Big)t\Big),  \quad   \Psi(t,x)= \exp\Big(\pm \sqrt{\frac{b_2}{d_{21}}}x-\Big(a_2+\frac{b_2d_2}{d_{21}}\Big)t\Big).\]
Conservation laws for systems of reaction-diffusion equations
without cross-diffusion were found in our earlier study
\cite{ch-king5}.

\section{Exact solutions and their properties} \label{sec-3}

From the applicability point of view, the systems listed in Cases
1--4, 7 and 9 of Table~\ref{tab1} preserve the main features of the
original SKT system \cite{sh-ka-te} because those involve
cross-diffusion terms in both equations  and  the reaction terms in
these systems  describe the competition of two populations provided
the  parameters $a_i,\  b_i$  and $c_i$ are nonnegative (for this
type of interaction, the linear terms must be nonnegative, while
quadratic terms must be nonpositive). Although  other systems
identified  in Table~\ref{tab1} can be also useful in some other
applications, we analyse only the afore-mentioned  systems  in what
follows.

In this section, our aim is to construct  exact solutions of the SKT
systems using their   conditional symmetries,
  and to identify  their  biological
 interpretations.  Each  $Q$-conditional symmetry
 operator listed in Cases 1--4, 7 and 9 of Table~\ref{tab1}  can be applied   for constructing an ansatz
reducing  the  relevant  SKT system
to  a system of ODEs. The standard algorithm (see  a  discussion and
examples for scalar PDEs   in Chapter 5 \cite{{bl-anco-10}},
nontrivial examples for reaction-diffusion systems being found in
Chapter 3 \cite{ch-dav-book}) was used in order to do this.
According to the algorithm, one should solve
 first-order PDE systems of the form
\be\nonumber\begin{array}{l}
 u_t+\xi u_x-\eta^1=0, \medskip\\
 v_t+\xi v_x-\eta^2=0, \end{array}\ee
 which are often called invariant surface conditions.  Here the functions $ \xi, \ \eta^1$   and $\eta^2$ are the known coefficients of the $Q$-conditional symmetry operator (\ref{2-29}).
  Thus, invariant surface conditions
  for each operator listed in Cases 1--4, 7 and 9 of Table~\ref{tab1}  were solved and the  ans\"atze obtained were used in order to obtain reduced systems of ODEs.
  The  results  are summarized in Table~\ref{tab3}, where $\varphi(x)$ and
$\psi(x)$ are new unknown  functions.

\begin{Remark}
Several ans\"atze in Table~\ref{tab3} can be simplified by the
renaming $\frac{\varphi(x)}{\psi(x)} \to \varphi(x)$; however, the
reduced systems of ODEs are more complicated  after introducing the
above  renaming because each ODE involves the  nonlinear term
$(\psi\varphi)''$. As a result, all the reduced equations would be
nonlinear.
\end{Remark}


\begin{table}
\caption{Ans\"atze and reduced systems of ODEs}\medskip
\label{tab3}       
\begin{center}
\begin{tabular}{|c|c|c|}
\hline Operator & Ansatz & Reduced system  \\
\hline

$Q^1$ &
$u=\left(a_2\psi(x)+\left(e^{a_2t}-1\right)\varphi(x)\right)e^{\gamma
t},$ & $\varphi''-b_2\varphi=0,$
 \\ $a_2\neq0$ &$v=\frac{a_2\varphi(x)e^{a_2t}}{a_2\psi(x)+\left(e^{a_2t}-1\right)\varphi(x)},$& $\psi''+
\left(a_1-\gamma\right)\psi-\left(1+c_1-b_2d_{12}\right)\varphi=0$
\\ &$\gamma=a_1+b_2-a_2(1+c_1-b_2d_{12})$& \\ \hline

$Q^1$ & $u=\left(\varphi(x)t+\psi(x)\right)e^{(a_1+b_2)t},$ &
$\varphi''-b_2\varphi=0$
 \\ $a_2=0$ &$v=\frac{\varphi(x)}{\varphi(x)\,t+\psi(x)}$&
 $\psi''-
b_2\psi-\left(1+c_1-b_2d_{12}\right)\varphi=0$
\\ \hline

$Q^2$ & $u=\psi(x),$ & $\varphi''-b_2\varphi=0$
 \\ &$v=\frac{\varphi(x)}{\psi(x)}\,e^{a_2t}$&
 $\psi''+\psi(a_1-b_1\psi)=0$
\\ \hline

$Q_0^3$ & $u=\psi(x),$ &
$\varphi''+\varphi\left(\frac{a_2}{\psi}-b_2\right)=0$
 \\ $a_2\neq0$ &$v=\frac{\varphi(x)}{\psi(x)}+\frac{f(x)}{\psi(x)}\,e^{a_2t}$&
 $\psi''+\psi(a_1-b_1\psi)-\frac{a_2\varphi}{\psi}=0$
\\ \hline

$Q_0^3$ & $u=\psi(x),$ & $\varphi''-b_2\varphi-\frac{f}{\psi}=0$
 \\ $a_2=0$ &$v=\frac{\varphi(x)}{\psi(x)}+\frac{tf(x)}{\psi(x)}$&
 $\psi''+\psi(a_1-b_1\psi)+\frac{f}{\psi}=0$
\\ \hline

$Q^4$ & $u=\psi(x)e^{\alpha t},$ & $\varphi''-b_2\varphi=0$
 \\  &$v=\frac{\varphi(x)}{\psi(x)}$&
 $\psi''+(a_1-\alpha)\psi-\left(c_1-b_2d_{12}\right)\varphi=0$
\\ \hline

$Q^5$ & $u=\psi(x)e^{\mu t}-d_2,$ & $\varphi''-b_2\varphi=0,$
 \\  &$v=\frac{\varphi(x)}{\psi(x)}e^{\mu t},$&
 $(d_2\varphi+d_1\psi^2)\psi''+d_2\left(\psi^2\right)'\left(\frac{\varphi}{\psi}\right)'+$
\\ &$\mu=a_2+b_2d_2$&$\mu\psi^3=0$ \\ \hline

$Q^8$ & $u=\psi(x)e^{\alpha t},$ & $\varphi''-b_2\varphi=0$
 \\  &$v=\frac{\varphi(x)}{\psi(x)}e^{a_2t}$&
 $\psi''+\left(a_1-\alpha\right)\psi=0$
\\ \hline

$Q_0^{11}$ & $u=\psi(x)e^{\alpha t},$ &
$\varphi''-b_2\varphi-\frac{f}{\psi}=0$
 \\ $\alpha\neq0$ &$v=\frac{\varphi(x)}{\psi(x)}+\frac{f(x)}{\psi(x)}\frac{e^{\alpha t}}{\alpha}$&
 $\psi''+\left(a_1-\alpha\right)\psi+\frac{f}{\psi}=0$
\\ \hline

$Q_0^{11}$ & $u=\psi(x),$ & $\varphi''-b_2\varphi-\frac{f}{\psi}=0$
 \\ $\alpha=0$ &$v=\frac{\varphi(x)}{\psi(x)}+\frac{f(x)}{\psi(x)}\,t$&
 $\psi''+a_1\psi+\frac{f}{\psi}=0$
\\ \hline

$Q^{11}$ & $u=\psi(x)e^{\beta t},$ & $\varphi''-b_2\varphi=0, \
\beta=a_1+b_2$
 \\ $\alpha_0\neq0$ &$v=\left(\beta+\alpha_0\left(e^{\beta t}-1\right)\right)\frac{\varphi(x)}{\psi(x)}+
 \frac{f(x)}{\psi(x)}-1$&
 $\psi''-b_2\psi+\alpha_0\beta\frac{\varphi}{\psi}=0$
\\ \hline
 \end{tabular}
 \end{center}
 \end{table}

By a simple analysis of Table~\ref{tab3}, one  notes that the
$Q$-conditional symmetries $Q^1, \ Q^4$ and $Q^8$ lead
 to  linear systems of ODEs even though the relevant systems of PDEs are nonlinear.
 Thus, these linear systems of ODEs  can be  integrated by a standard technique.
 As a result,       multiparameter families of solutions of the nonlinear PDEs in question
 can be readily  constructed. In particular, all exact solutions, which are obtainable using
the  conditional symmetry $Q^1$ are  presented below (see
Subsection~\ref{subsec-3.1}).

 All other $Q$-conditional symmetries listed in Table~\ref{tab3} lead to nonlinear systems of ODEs.
 Although such systems are simpler objects for analysis than the initial systems of PDEs, a general
  theory for their integration does not exist. So, we were able to
  derive only some particular results (see Subsection~\ref{subsec-3.2}).

\subsection{ Case 1 of Table~\ref{tab1}} \label{subsec-3.1}

 Let us consider the nonlinear  system  presented in Case 1 of Table~\ref{tab1}
\be\label{3-0}\begin{array}{l}
 u_t = [(1+d_{12}v)u]_{xx}+u\left(a_1-c_1v\right), \medskip\\
 v_t = [uv]_{xx}+v(a_2-b_2u-v),  \end{array}\ee
 where all parameters are arbitrary constants.
It can be shown  that the SKT system (\ref{3-0}) is a canonical form
of a more general SKT system. In fact,  the SKT system
 \be\label{3-0*}\begin{array}{l}
 u_t = [(d_1+d_{12}v)u]_{xx}+u\left(a_1-c_1v\right), \medskip\\
 v_t = d_{21}[uv]_{xx}+v(a_2-b_2u-c_2v) \end{array}\ee
is reducible to  (\ref{3-0}) by applying the equivalence (scale)
transformation
\[t \rightarrow d_1t, \ u \rightarrow \frac{d_1}{d_{21}}u, \ v\rightarrow \frac{d_1}{c_2} v,\]
and renaming \[d_{12} \rightarrow  c_2d_{12}, \ a_1\rightarrow
a_1d_1, \ a_2\rightarrow a_2d_1, \ c_1\rightarrow c_1 c_2, \
b_2\rightarrow b_2d_{21}.\] Hereafter  $d_1d_{12}\not=0$ because
restrictions  (\ref{2-2}) are assumed and  $c_2\not=0$ (otherwise
(\ref{3-0*}) is not reducible to (\ref{3-0})).

 Now our aim is to construct exact solutions of the SKT system (\ref{3-0})
using its $Q$-conditional symmetry operator $Q^1$. Depending on the
parameter $a_2$ (see the first two cases of Table~\ref{tab3}), we
have the ansatz
\begin{equation}\label{3-2}\begin{array}{l}
u(t,x)=\left(a_2\psi(x)+\left(e^{a_2t}-1\right)\varphi(x)\right)e^{\gamma t}, \ \gamma=a_1+b_2-a_2(1+c_1-b_2d_{12}),  \medskip \\
v(t,x)=\frac{a_2\varphi(x)e^{a_2t}}{a_2\psi(x)+\left(e^{a_2t}-1\right)\varphi(x)},\end{array}\end{equation}
if $a_2\neq0$, and the ansatz
\begin{equation}\label{3-3}\begin{array}{l}
u(t,x)=\left(\varphi(x)t+\psi(x)\right)e^{(a_1+b_2)t},\medskip
\\ v(t,x)=\frac{\varphi(x)}{\varphi(x)\,t+\psi(x)},\end{array}\end{equation} if $a_2=0$.

The ODE system corresponding to ansatz (\ref{3-2}) has the form
\begin{equation}\label{3-4}\begin{array}{l}
\varphi''-b_2\varphi=0,\medskip
\\ \psi''+
(a_1-\gamma)\psi-\left(1+c_1-b_2d_{12}\right)\varphi=0.\end{array}\end{equation}

It turns out that ansatz  (\ref{3-3}) again  leads to the ODE system
(\ref{3-4}) but the  expression for $\gamma$ simplifies to the form
$\gamma=a_1+b_2$ because $a_2=0$.


In contrast to   the SKT system  in question, which is nonlinear,
the reduced system of ODEs is a linear one and can be easily
integrated. The general solution of the first equation of
(\ref{3-4}) has the form \be\label{3-7}\varphi(x)=\left\{
\begin{array}{l}
C_1\,e^{\sqrt b_2 x}+C_2\,e^{-\sqrt b_2 x}, \ \mbox{if} \ b_2>0,
\medskip \\  \medskip
C_1\cos(\sqrt{-b_2}\,x)+C_2\sin(\sqrt{-b_2}\,x), \ \mbox{if} \ b_2<0,\\
C_1+C_2x, \  \mbox{if} \  b_2=0.\\
  \end{array} \right. \ee
Substituting the function $\varphi$ into the
 second equation of the ODE system (\ref{3-4}), one easily obtains
  \be\label{3-8}\psi(x)=\left\{
\begin{array}{l}
C_3e^{\kappa\,x}+C_4e^{-\kappa\,x}+\frac{\varphi(x)}{a_2}, \ \mbox{if} \ a_2(1+c_1-b_2d_{12})-b_2<0,  \medskip \\
C_3\cos(\kappa\,x)+C_4\sin(\kappa\,x)+\frac{\varphi(x)}{a_2}, \ \mbox{if} \ a_2(1+c_1-b_2d_{12})-b_2>0, \medskip\\
 C_3+C_4x+\frac{\varphi(x)}{a_2}, \  \mbox{if} \  a_2(1+c_1-b_2d_{12})-b_2=0,\\
  \end{array} \right.\ee in the case $a_2\neq0.$ Here
  $\kappa=\sqrt{|a_1-\gamma|} =\sqrt{|a_2(1+c_1-b_2d_{12})-b_2|}$ and $C_i \, (i=1,\dots,4)$  are arbitrary constants.

  In the case $a_2=0$, the function $\psi(x)$ has the form
 \[\psi(x)=\left\{
\begin{array}{l}
C_3\,e^{\sqrt b_2 x}+
 C_4\,e^{-\sqrt b_2 x}+\frac{(1+c_1-b_2d_{12})x}{2b_2}\varphi',   \ \mbox{if} \ b_2>0,  \medskip \\
C_3\cos(\sqrt{-b_2}\,x)+
 C_4\sin(\sqrt{-b_2}\,x)+\frac{(1+c_1-b_2d_{12})x}{2b_2}\varphi',  \ \mbox{if} \ b_2<0, \medskip\\
C_3+C_4x+\frac{(1+c_1)x^2}{6}(3C_1+C_2x),  \ \mbox{if} \ b_2=0.\\
  \end{array} \right.\]

Substituting the functions $\varphi$ and  $\psi$ from
the above formulae  into ans\"atze (\ref{3-2}) and (\ref{3-3}),  one
immediately  obtains four-parameter families of exact solutions for
the nonlinear  system (\ref{3-0}).

Now we turn to the biological sense of the nonlinear system
(\ref{3-0}), which is equivalent to (\ref{3-0*}). One realizes that
system (\ref{3-0}) is nothing else but the SKT system  (\ref{2-2})
with only two restrictions $b_1=d_2=0$. This  means from the
biological point of view that the Malthusian law is assumed for the
growth of  the first species  and
 the standard diffusivity of the second species is negligible comparing with the cross-diffusivity.
Thus, the system is an important  particular cases of the SKT system
(\ref{2-1}) if the following parameter restrictions take place:
   $a_1 $ and  $a_2$ are arbitrary nonnegative constants,
   $c_1 $ and  $b_2$ are   positive in order to have competition  between two populations of species.
 The parameter $d_{12}\not=0$ can have any sign provided the diffusivity $d_1+d_{12}v \geq 0$
 for all relevant $v$.

 Let us assume that the above  parameters  satisfy the  inequality $\gamma=a_1+b_2-a_2(1+c_1-b_2d_{12})<a_1$, i.e., $b_2-a_2(1+c_1-b_2d_{12})<0.$
 Then the relevant family of  exact solutions of the SKT
system (\ref{3-0}) can be identified as follows (see the first line
in (\ref{3-7}) and the second line in (\ref{3-8}))
\begin{equation}\label{3-10}\begin{array}{l}
u(t,x)=e^{(\gamma+a_2)t}\Big(C_1e^{\sqrt b_2 x} + C_2e^{-\sqrt b_2 x} +C_3e^{-a_2t}\sin(\sqrt{a_1-\gamma}\,(x+x_0))\Big),\medskip\\
v(t,x)=a_2\,\frac{C_1e^{\sqrt b_2 x} + C_2e^{-\sqrt b_2
x}}{C_1e^{\sqrt b_2 x} + C_2e^{-\sqrt b_2 x}
+C_3e^{-a_2t}\sin(\sqrt{a_1-\gamma}\,(x+x_0))}.
\end{array}\end{equation}
In the above formulae, a linear combination of elementary
trigonometric functions has been  replaced by the single sine
function with the parameter $x_0$ instead of $C_4$.

Let us assume that the competition of two populations occurs in the
domain
 \[\Omega=\lf\{ (t,x) \in (0,+ \infty
)\times (a,b)\rg\}, \quad a<b\in \mathbb{R}. \]
Setting  $C_3\not=0$ (otherwise the component $v(t,x)=a_2$), we can identify the restrictions on  $C_1$ and  $C_2$ that  guarantee boundedness  and nonnegativity of the both components. 
 The restrictions are
\[  |C_3|< 2\sqrt{C_1C_2}, \ C_1>0,  \  C_2>0. \]
 In the case $C_1C_2=0$, the corresponding  restrictions are
\[  |C_3|< C_1, \  a\geq0  \ \mbox{or} \  |C_3|< C_2, \  b\leq0. \]

Moreover, depending on relations between $\gamma$  and $a_2$,  exact
solutions  of the form (\ref{3-10}) can describe three different
scenarios of  species  competition.

\medskip
If \quad  $\gamma<-a_2 \Leftrightarrow
d_{12}<\frac{a_2c_1-a_1-b_2}{a_2b_2}$ \quad  then \quad  $(u,\,v)
\rightarrow (0,\,a_2) \quad \mbox{as} \quad  t  \rightarrow +\infty.
$ \medskip

 This implies   the complete extinction of the species  $u$, while  the species  $v$ survives. The reason for such a scenario can be the following:
 the cross-diffusion coefficient $d_{12}$ is  small, so that  the diffusion pressure $d_{12}v$ caused by the species  $v$  is too small comparing to the reaction term $c_1vu$, which has the opposite sign (see the first equation in (\ref{3-0})).  Notably   $(0,\,a_2)$ is a steady-state point of   the SKT system (\ref{3-0}).

\medskip
 If \quad  $\gamma>-a_2 \Leftrightarrow d_{12}> \frac{a_2c_1-a_1-b_2}{a_2b_2}$  \quad
 then \quad
 $  (u,\,v)  \rightarrow  (+\infty,\,a_2) \quad \mbox{as} \quad  t  \rightarrow +\infty.
$ \medskip

 This represents   unbounded growth  of the species  $u$, while  the species  $v$ again survives.
 Such a scenario can again be caused by the cross-diffusion coefficient $d_{12}$  because that is large in this case. However, taking into account the above restriction for the existence of  exact solutions  of the form  (\ref{3-10}), the required  interval
 $\Big( \frac{a_2c_1-a_1-b_2}{a_2b_2}, \frac{a_2c_1+a_2-b_2}{a_2b_2}\Big)$ for  $d_{12}$ can be found.

Finally, there is a special value for the cross-diffusion
coefficient $d_{12}=d^{\,0}_{12}=\frac{a_2c_1-a_1-b_2}{a_2b_2}$
(meaning $\gamma=-a_2 $ )  when  coexistence of the  two populations
takes place:
 \[ (u,\,v)  \rightarrow  \left(C_1e^{\sqrt b_2 x} + C_2e^{-\sqrt b_2 x},\,a_2\right)
 \quad \mbox{as} \quad  t  \rightarrow +\infty. \]

\begin{figure}[h!]
\begin{center}
\includegraphics[width=7.5cm]{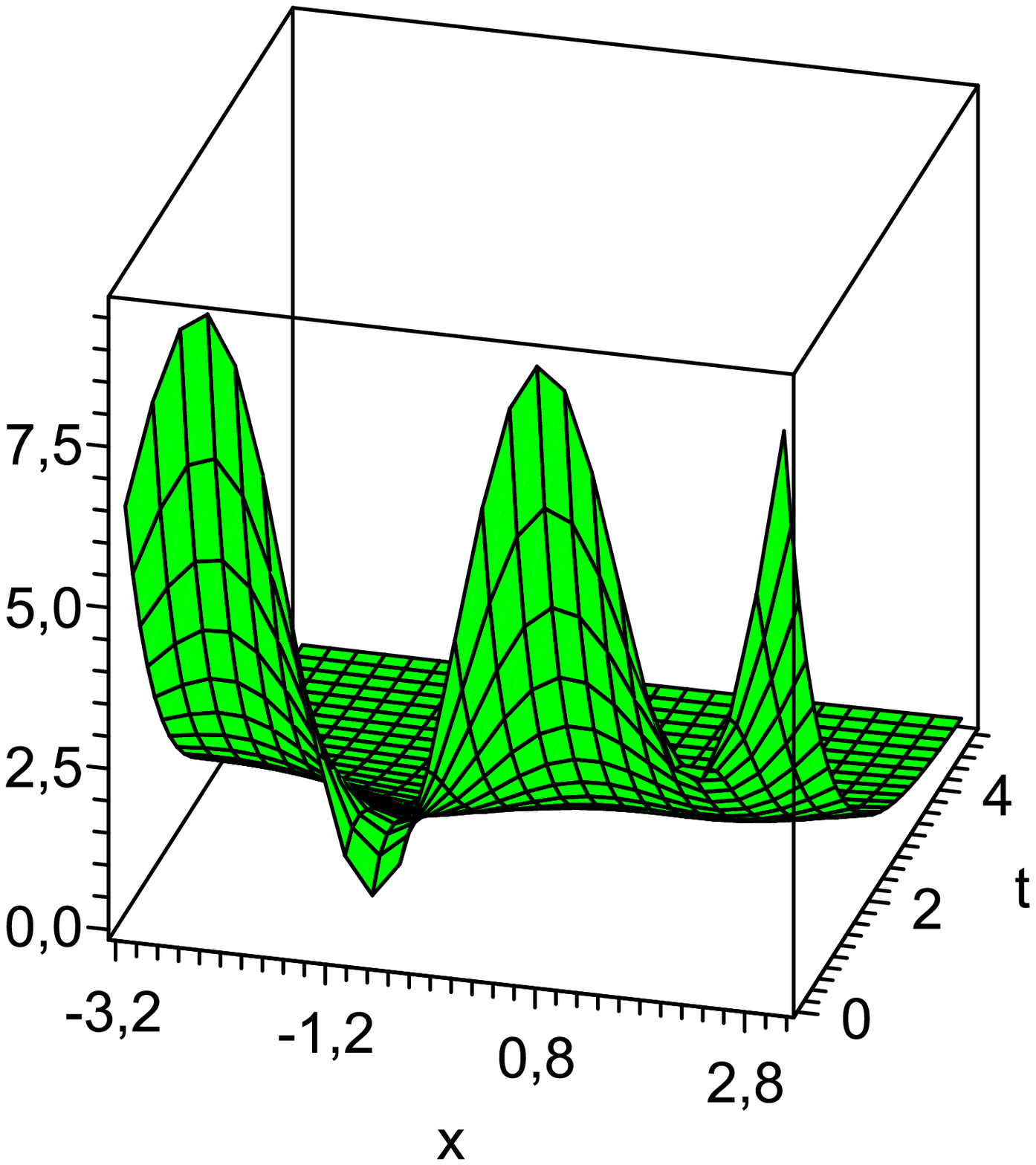}
\includegraphics[width=7.5cm]{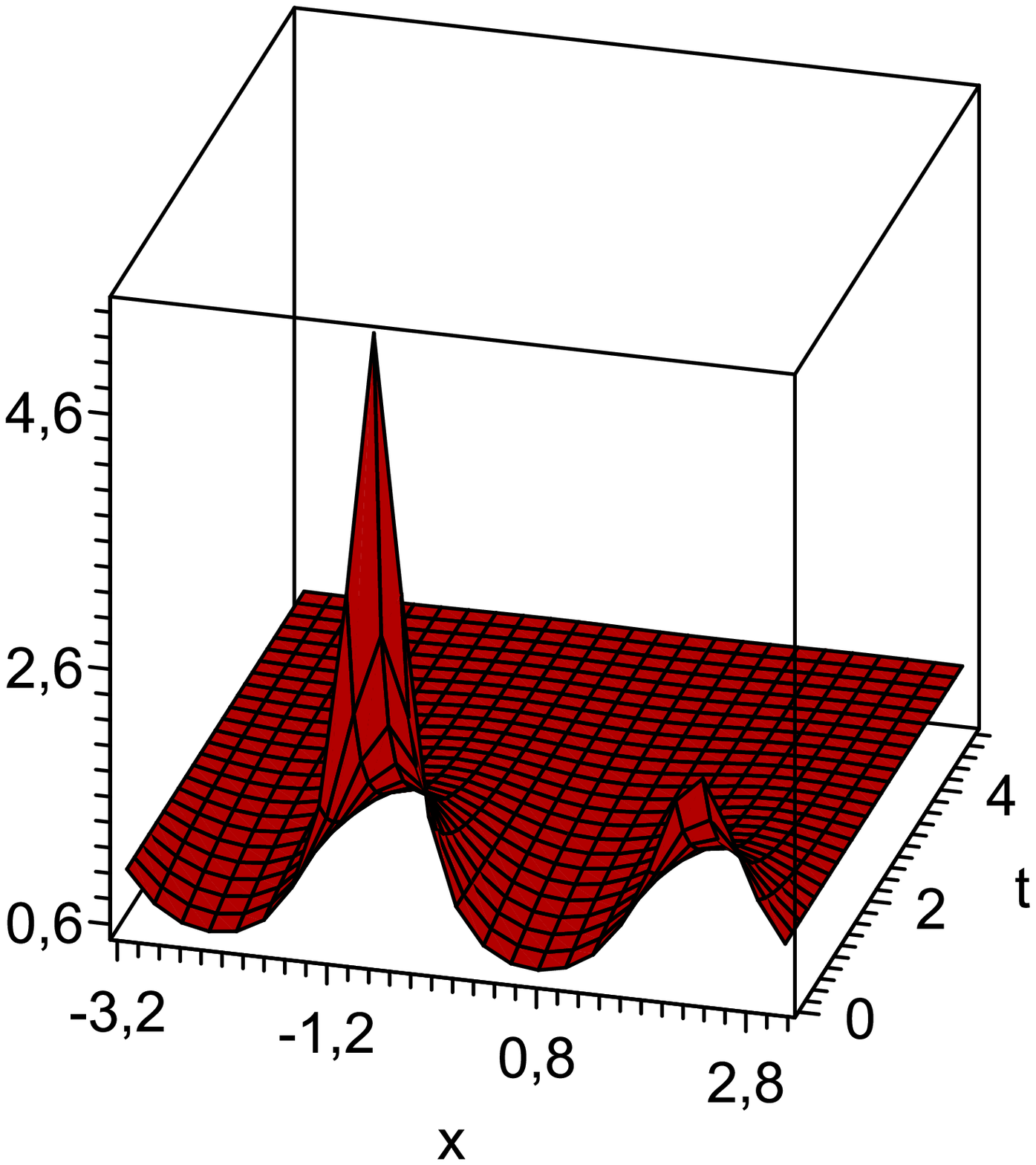}
\end{center}
\caption{Surfaces representing the   $u$ (green) and $v$ (red)
components  of solution (\ref{3-10})
 of the SKT system (\ref{3-0}) with the parameters $a_1=2, \ a_2=1, \
 \gamma=-2, \ b_2=\frac{1}{10}, \ d_{12}=10c_1-31, \
 C_1=3, \ C_2=2, \ C_3=4, \ x_0=0.$} \label{f1}
\end{figure}

\begin{figure}[h!]
\begin{center}
\includegraphics[width=7.5cm]{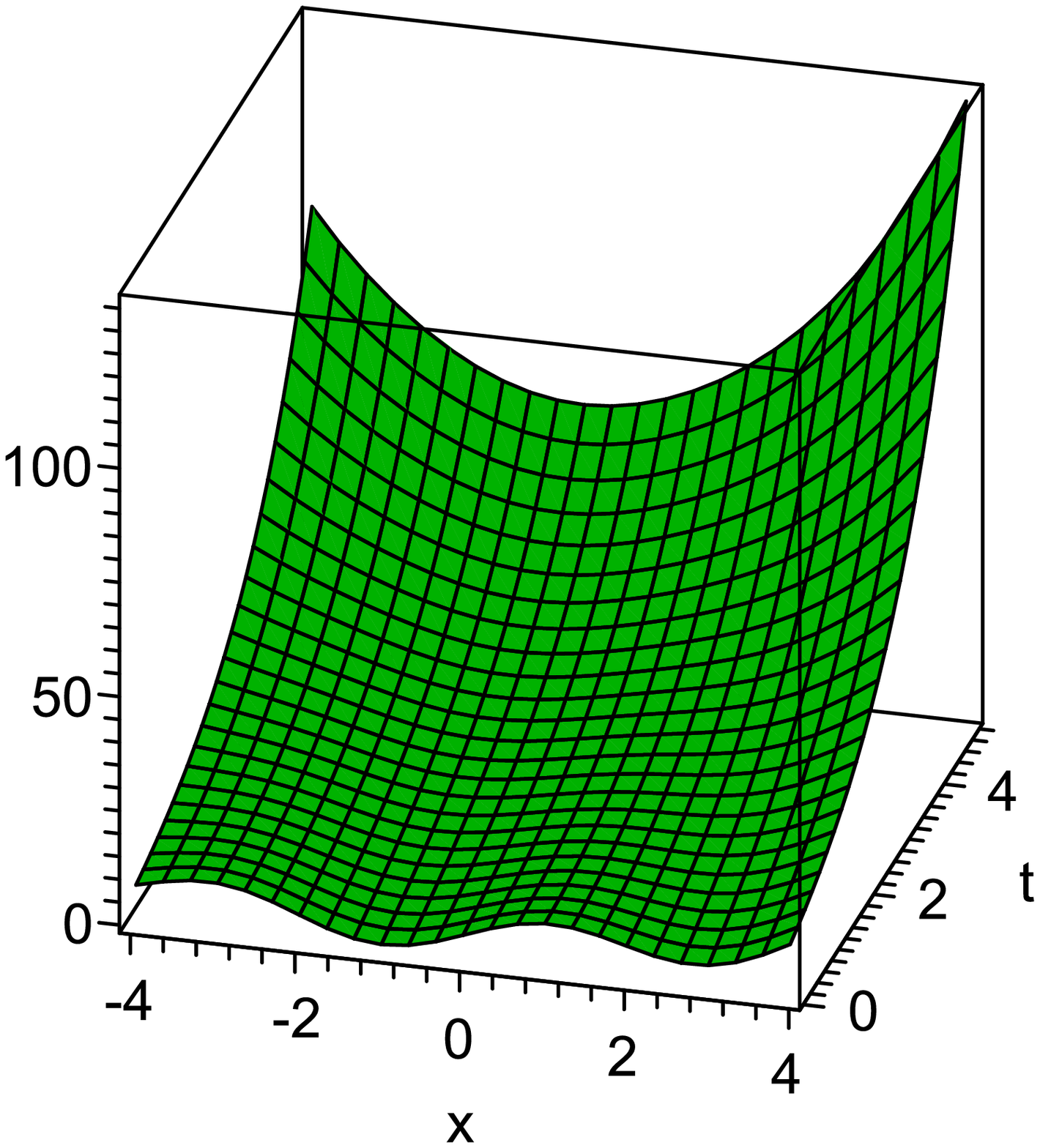}
\includegraphics[width=7.5cm]{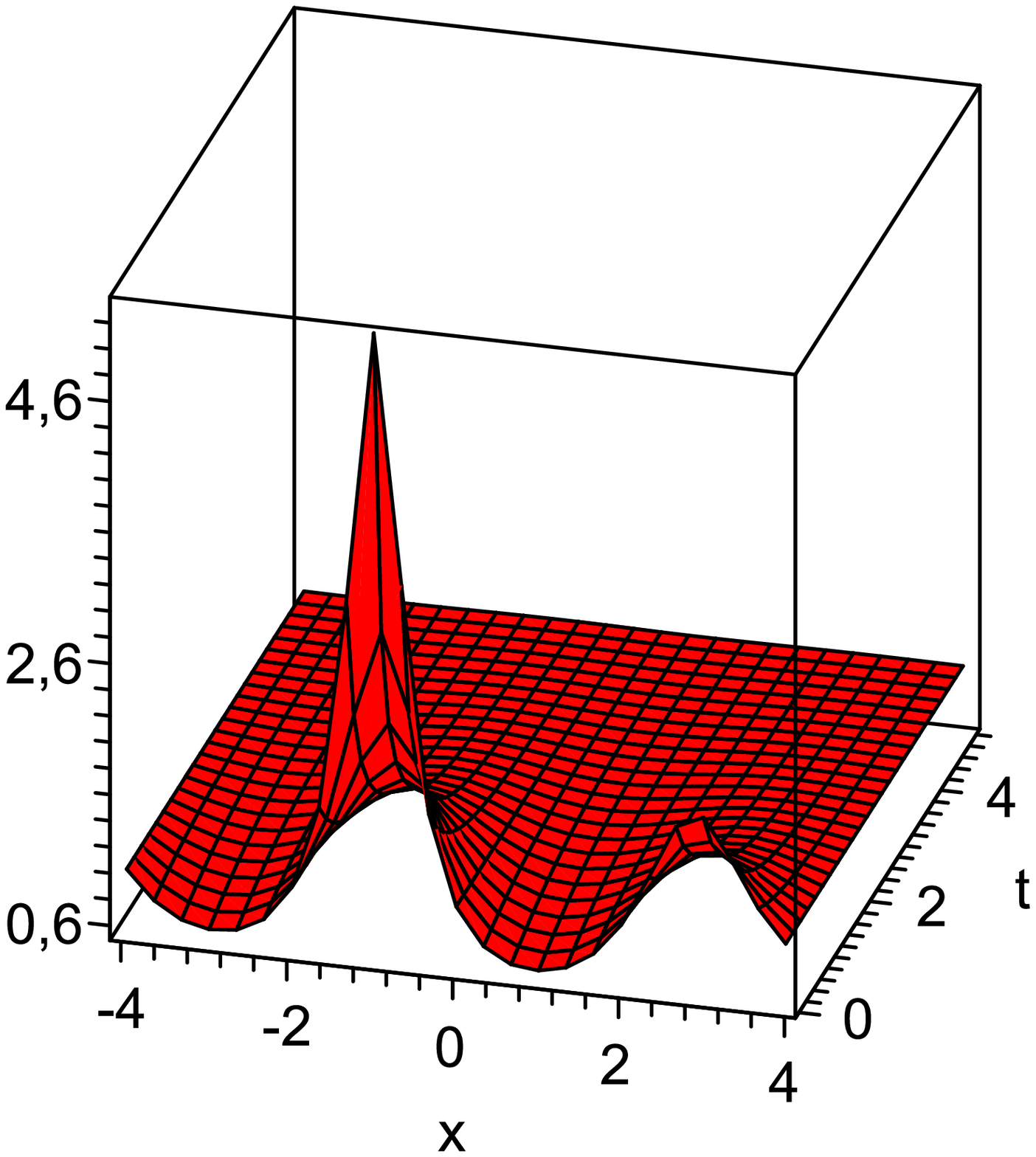}
\end{center}
\caption{Surfaces representing the $u$ (green) and $v$ (red)
components  of solution (\ref{3-10})
 of the SKT system (\ref{3-0}) with the parameters $a_1=2, \ a_2=1, \
 \gamma=-\frac{1}{2}, \ b_2=\frac{1}{10}, \ d_{12}=10c_1-16, \
 C_1=3, \ C_2=2, \ C_3=4, \ x_0=0.$} \label{f2}
\end{figure}

\begin{figure}[h!]
\begin{center}
\includegraphics[width=7.5cm]{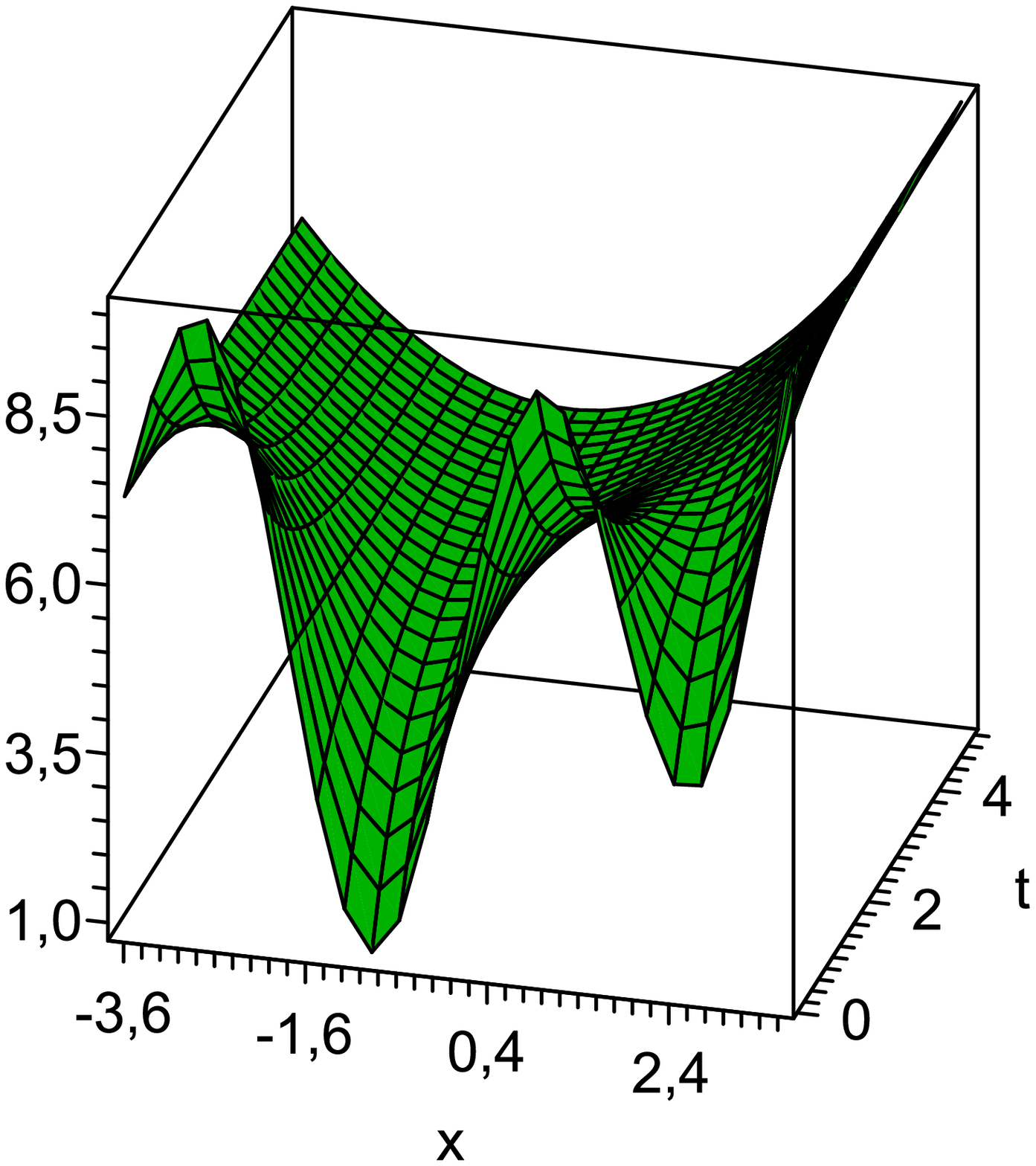}
\includegraphics[width=7.5cm]{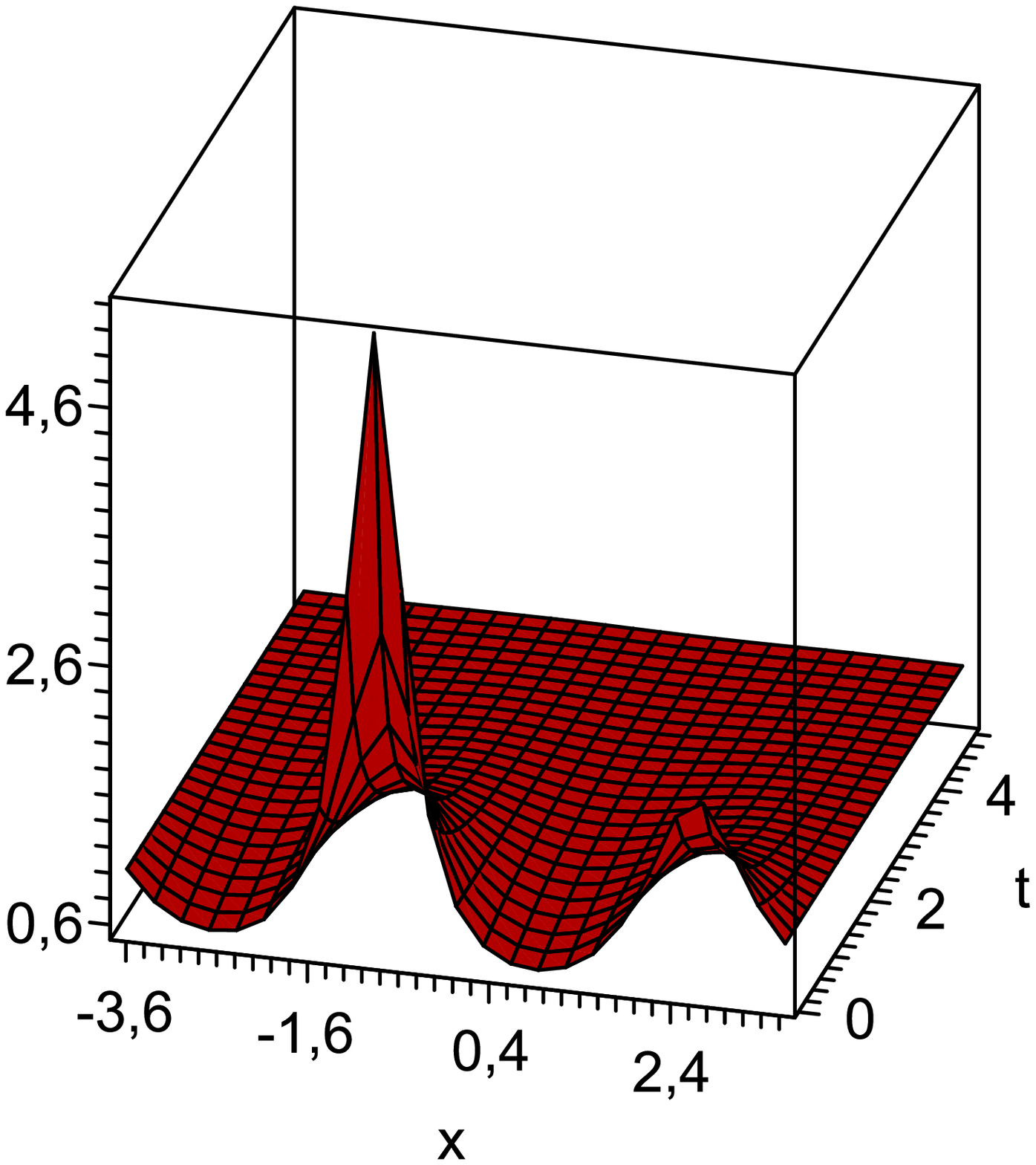}
\end{center}
\caption{Surfaces representing the $u$ (green) and $v$ (red)
components  of solution (\ref{3-10})
 of the SKT system (\ref{3-0}) with the parameters $a_1=2, \ a_2=1, \
 \gamma=-1, \ b_2=\frac{1}{10}, \ d_{12}=10c_1-21, \
 C_1=3, \ C_2=2, \ C_3=4, \ x_0=0.$} \label{f3}
\end{figure}

 Figures\,\ref{f1}, \ref{f2} and \ref{f3}
demonstrate three essentially different scenarios of competition
between the  species $u$ and $v$  that  are described by the exact
solution (\ref{3-10}).


 It can be also noted that the exact solution  (\ref{3-10}) with $d^{\,0}_{12}, \, x_0=0, $ and correctly-specified  $C_1$ and  $C_2$ solves the boundary value
 problem consisting of  the governing Eqs.\,(\ref{3-0}) and the   Dirichlet  conditions
\[\begin{array}{l} x=a: \,  u=U_a,  \, v=a_2, \\
x=b: \,  u=U_b,  \, v=a_2 \end{array} \] in the domain $\Omega$ with
the space interval $(a,b)= \frac{\pi}{\sqrt{a_1+a_2}}(k_1,k_2), \
k_1<k_2 \in \mathbb{Z}. $ Here $U_a$  and $ U_b$  are arbitrary
given constants.

\subsection{ Case 9 of Table~\ref{tab1}} \label{subsec-3.2}

Let us construct exact solutions of the  system  arising in  Case 9
of Table~\ref{tab1} with some positive  parameters $a_1$ and $b_2$.
Note that this system can still be regarded as a simplified SKT
system  because one
 preserves cross-diffusion terms in both equations and the reaction terms reflecting the competition between species, i.e.  the main fetures of the initial system are preserved.
 Using the equivalence (scale)
transformation $t\rightarrow \frac{1}{b_2}\,t, \
x\rightarrow\frac{1}{\sqrt{b_2}}\,x$ and renaming $a_1\rightarrow
a_1b_2$, one can again to simplify the system in question to the
form
 \be\label{2-14}\begin{array}{l}
  u_t = [(1+v)u]_{xx}+u\left(a_1-v\right), \medskip\\
   v_t = [uv]_{xx}-uv. \end{array}\ee
   As one may note,   the SKT system (\ref{2-14}) possesses a rich symmetry because it is invariant 
 with respect to  two operators of $Q$-conditional symmetry,  $Q_0^{11}$ and $Q^{11}$. Moreover, both operators contains several arbitrary constants  as parameters.

Ans\"atze corresponding to the operator $Q_0^{11}$ (see
Table~\ref{tab3}) have the forms
\begin{equation}\label{3-16}\begin{array}{l}
\begin{array}{l}u(t,x)=\psi(x)e^{\alpha t},
\\ v(t,x)=\frac{\varphi(x)}{\psi(x)}+\frac{f(x)}{\psi(x)}\frac{e^{\alpha
t}}{\alpha},\end{array} \hskip1cm \mbox{if} \ \alpha\neq0, \medskip \\
\begin{array}{l} u(t,x)=\psi(x),
\\
v(t,x)=\frac{\varphi(x)}{\psi(x)}+\frac{f(x)}{\psi(x)}\,t,\end{array}
\hskip1.3cm \mbox{if} \  \alpha=0,\end{array}
\end{equation}
while the operator $Q^{11}$ produces   the ansatz
\begin{equation}\label{3-17}\begin{array}{l}
u(t,x)=\psi(x)e^{(1+a_1) t},
\\ v(t,x)=\left(1+a_1+\alpha_0\left(e^{(a_1+1)t}-1\right)\right)\frac{\varphi(x)}{\psi(x)}+\frac{f(x)}{\psi(x)}-1,
 \ \alpha_0\neq0.
 \end{array}
\end{equation}
In formulae (\ref{3-16}) and (\ref{3-17}), $\varphi$ and $\psi$ are
to-be-determined functions, while
$f(x)=\alpha_1e^{x}+\alpha_2e^{-x}$ (see  (\ref{2-45})  with
$b_2=1$).

The ODE system corresponding to the first ansatz from (\ref{3-16})
has the form (see Table~\ref{tab3})
\begin{equation}\label{3-18}\begin{array}{l}
\varphi''-\varphi-\frac{f}{\psi}=0,\medskip
\\ \psi''+(a_1-\alpha)\psi+\frac{f}{\psi}=0.\end{array}\end{equation}
The second ansatz from (\ref{3-16}) leads to the ODE system
(\ref{3-18}) with $\alpha=0$.

The ODE system corresponding to ansatz (\ref{3-17}) has the form
\begin{equation}\label{3-19}\begin{array}{l}
\varphi''-\varphi=0,\medskip
\\ \psi\psi''-\psi^2+\alpha_0(1+a_1)\varphi=0.\end{array}\end{equation}

The systems (\ref{3-18}) and (\ref{3-19})  are both nonlinear,
consisting of  second order ODEs. To the best of our knowledge, the
general solutions of these systems are unknown; therefore  only
particular solutions were identified by  applying  additional
restrictions.


 In order to
construct particular  solutions of the ODE system (\ref{3-18}), we
examine two cases: $f=0$  (i.e., $\alpha_1=\alpha_2=0$) and
$f\not=0$.

In the case $f=0$,  system (\ref{3-18}) can be easily integrated and
we obtain the function $\psi(x)$ as follow
 \[\psi(x)=\left\{
\begin{array}{l}
C_1e^{\sqrt{\alpha-a_1}\,x}+C_2e^{-\sqrt{\alpha-a_1}\,x}, \ \mbox{if} \ \alpha-a_1>0,  \medskip \\
C_1\cos\left(\sqrt{a_1-\alpha}\,x\right)+C_2\sin\left(\sqrt{a_1-\alpha}\,x\right), \ \mbox{if} \ \alpha-a_1<0, \medskip\\
 C_1+C_2x, \  \mbox{if} \  \alpha-a_1=0,\\
  \end{array} \right.\]
while the function $\varphi(x)=C_3e^{x}+C_4e^{-x}$. Thus, the
relevant families of exact solutions of the SKT system in question
can easily be  written down. In order to avoid exact solutions with
unbounded growth in time, we set $\alpha<0$, so the  family of exact
solutions of the SKT system (\ref{2-14})
\begin{equation}\label{3-19*}\begin{array}{l} u(t,x)=\Big(C_1\cos\left(\sqrt{a_1-\alpha}\,x\right)+
C_2\sin\left(\sqrt{a_1-\alpha}\,x\right)\Big)e^{\alpha t},\medskip\\
v(t,x)=\frac{C_3e^{x}+C_4e^{-x}}{C_1\cos\left(\sqrt{a_1-\alpha}\,x\right)+C_2\sin\left(\sqrt{a_1-\alpha}\,x\right)}
\end{array}\end{equation}
is constructed using the first ansatz in (\ref{3-16}).

To find exact solutions of system (\ref{3-18}) in the case $f\neq0$,
we assume that the function $\psi(x)=C_1 e^{\gamma x}$ (here $C_1$
and $\gamma$ are arbitrary nonzero constants at the moment).
Substituting the function $\psi$ into the second equation of system
(\ref{3-18}), we find the set of the specified constants
 \[\gamma=-\frac{1}{2}, \ \alpha_1=0, \
\alpha_2=C_1^2\left(\alpha-a_1-\frac{1}{4}\right),\] when
$\psi(x)=C_1 e^{- x/2}$ is indeed an exact solution.
 Integrating the
first equation of system (\ref{3-18}), we obtain
\[\varphi(x)=-\frac{4}{3}C_1\left(\alpha-a_1-\frac{1}{4}\right) e^{-\frac{1}{2}\,x}+
C_2e^{-x}+C_3e^{x}.\] Substituting the functions $\varphi_1$ and
$\varphi_2$ into the first ansatz in  (\ref{3-16}), we arrive at the
exact solution
 \be\label{3-20}\begin{array}{l}
  u(t,x)=C_1e^{\alpha t-\frac{1}{2}\,x}, \medskip\\
   v(t,x)= \frac{1+4(a_1-\alpha)}{3} -\frac{1+4(a_1-\alpha)}{4\alpha}C_1e^{\alpha t-\frac{1}{2}\,x} + C_2e^{-\frac{1}{2}\,x}+
   C_3e^{\frac{3}{2}\,x}
   \end{array}\ee
of the SKT system (\ref{2-14}). This is a  four-parameter family of
solutions and we want to find a spatial domain in which both
components are bounded and nonnegative. Obviously, the   domain
could be  a semi-infinite   interval or a finite interval.
Let us consider the domain with  the semi-infinite   interval
 \[\Omega=\lf\{ (t,x) \in (0,+ \infty
)\times (0,+ \infty )\rg\}.\] It can easily be  shown that both
components of  solution  (\ref{3-20}) are bounded and nonnegative in
$\Omega$  provided  the following  restrictions take place:
\be\label{3-21} \alpha<0, \ C_1>0, \ C_3=0, \
C_2+\frac{1+4(a_1-\alpha)}{3}>0 \ee (in the case of nonnegative
$C_2$ the last inequality is fulfilled automatically).
 Moreover,  the exact solution (\ref{3-20}) with restrictions (\ref{3-21})
 has the time asymptotic behaviour
  \[ \lf(u,\ v\rg)\rightarrow \lf(0, \ \frac{1+4(a_1-\alpha)}{3}+C_2e^{-\frac{1}{2}\,x}\rg), \quad \mbox{as} \quad
t\rightarrow +\infty.\]
 This means one can  describe
competition between  the two species
in which the species $v$ eventually dominates  while the  species
$u$ dies. Note that solution  (\ref{3-20}) with $C_2=C_3=0$ is a
plane wave solution.

Particular solutions of the ODE system (\ref{3-19}) are obtainable
in a quite similar way as those for (\ref{3-18}). However,
application of  ansatz (\ref{3-17}) to the solutions obtained leads
to exact solutions of the SKT system  (\ref{2-14}) with $u$  and $v$
becoming unbounded as $t\rightarrow +\infty $,
  which is rather unrealistic. In the special
case $\varphi=0$, the ODE system (\ref{3-19}) leads to
 the family of exact solution of the SKT system (\ref{2-14})
 \[\begin{array}{l}
  u(t,x)=  (C_1e^{x}+  C_2e^{-x})e^{(1+a_1)t}, \medskip\\
   v(t,x)= \frac{\alpha_1e^{x}+  \alpha_2e^{-x}}{C_1e^{x}+  C_2e^{-x}} -1.
   \end{array}\]
   Depending on the parameter values, essentially different forms of the component $v$ arise,
   for example, $v=const, \,  v=\tanh x,  \,  v=\coth x$  can be  obtained.

Finally, we note that the SKT system (\ref{2-14})
can be considered as a canonical form of the more general system
involving 5 parameters \be\label{3-22}\begin{array}{l}
  u_t = [(d_1+d_{12}v)u]_{xx}+u\left(a-b\,v\right), \medskip\\
   v_t = d_{21}[uv]_{xx}-\frac{b\,d_{21}}{d_{12}}uv, \quad  d_{12}\not=0. \end{array}\ee
In fact, system (\ref{3-22}) is reducible to  (\ref{2-14}) by
applying the equivalence transformation
\[t\rightarrow \frac{bd_1}{d_{12}}\,t, \ x\rightarrow \sqrt{\frac{b}{d_{12}}}\,x, \ u\rightarrow \frac{d_1}{d_{21}}\,u,
 \ v\rightarrow
\frac{d_1}{d_{12}}\,v, \ a\rightarrow \frac{bd_1}{d_{12}}\,a_1.\]

\section{Alternative application of  system (\ref{2-1})} \label{sec-4}
Further motivation for the current study can be provided by noting
that some of the systems in question
 can be interpreted as relevant to polymerisation rather than ecology.

Let $u$ and $v$ be the concentrations of two distinct species of
monomers and $w$ be that of their heterodimer. The simplest
applicable reaction-diffusion model then takes the form
\be\label{B-1}
 \ba
  u_t = d_1u_{xx}+k_-w-k_+uv-a_1u,\\
 v_t = d_2v_{xx} + k_-w-k_+uv-a_2v,\\
 w_t = d_3w_{xx} - k_-w + k_+uv-a_3w,
 \ea
  \ee
where the positive constants $d_i$  and $a_i \, (i=1,2,3)$ are
associated diffusivities and degradation rates, respectively, $k_+$
specifies the  dimerisation rate and  $k_-$ that of fragmentation.

In the physically  relevant limits in which   $k_-$ and $k_+$ are
sufficiently large,
 the above three-component  system simplifies to a two-component system
with
  \[ w= Kuv, \quad  K= \frac{k_+}{k_-}, \]
 at leading order,  where $K$ is the relevant equilibrium constant.
  Now  adding the first and third equations, and second  and third equations from (\ref{B-1})
  under the additional assumptions $|w_t|\ll |u_t|$ and  $|w_t|\ll |v_t|$,
  which require the less-physically appropriate assumptions that $d_3\gg d_1$ and $d_3\gg d_2$
  with $K$ small, one arrives at the system

  \be\label{B-2}
 \ba
  u_t = [(d_1+d_{3}Kv)u]_{xx} -u(a_1 + a_3Kv),\\
 v_t = [(d_2+d_{3}Ku)v]_{xx}-v(a_2 + a_3Ku).
 \ea
  \ee
  Obviously, the two-component system (\ref{B-2}) is a particular case of the simplified  SKT system (\ref{2-1}),  though it   involves  linear terms $ a_1u$ and $a_2v$ with  signs opposite from those typically adopted in the ecological content.
  Now one realizes that system (\ref{B-2}) coincides (up to scale transformations) with that listed in Case 5 of Table~\ref{tab1}.
Thus, the latter admits the conditional symmetry operator~$Q^6$.

  System (\ref{B-2}) is reducible to the   system
 \be\label{B-9}
 \ba
  u_t = [(d_1+v)u]_{xx} -u(a_1 + v),\\
 v_t = [(d_2+u)v]_{xx}-v(a_2 + u),
 \ea
  \ee
by applying the scale transformation \be \label{B-10}x\rightarrow
\sqrt{\frac{d_3}{a_3}}\,x, \ u\rightarrow \frac{1}{a_3K}\,u,
 \ v\rightarrow
\frac{1}{a_3K}\,v, \ d_1\rightarrow \frac{d_3}{a_3}\,d_1, \
d_2\rightarrow \frac{d_3}{a_3}\,d_2.\ee
 Setting in (\ref{B-9}) $a_1=\frac{d_1^2}{d_1+d_2}, \
a_2=\frac{d_2^2}{d_1+d_2}$ and using operator $Q^6$ with $b_2=1$,
one obtains  the ansatz
\be\label{B-3}\ba u(t,x)=\varphi(x)\exp\lf(\frac{d_1d_2}{d_1+d_2}\,t\rg)-\frac{d_2^2}{d_1+d_2},\medskip\\
v(t,x)=\psi(x)\exp\lf(\frac{d_1d_2}{d_1+d_2}\,t\rg)-\frac{d_1^2}{d_1+d_2},\ea\ee
and the reduced ODE system
\be\label{B-4}\ba\lf(d_1\varphi+d_2\psi\rg)\varphi''+2d_2\varphi'\psi'-d_1\varphi^2=0,\medskip\\
\lf(d_1\varphi+d_2\psi\rg)\psi''+2d_1\varphi'\psi'-d_2\psi^2=0.\ea\ee
Although the latter is a system of nonlinear second-order ODEs, it
turns out  that its general solution can be derived. In fact,
writing down   a linear combination of the above ODEs with
coefficients $d_1$ and $-d_2$, we arrive at the equation
\be\nonumber
\lf(d_1\varphi+d_2\psi\rg)\lf(d_1\varphi''-d_1\varphi-d_2\psi''+d_2\psi\rg)=0.\ee
The case $d_1\varphi+d_2\psi=0$ leads to the equation
$2{\psi'}^2+\psi^2=0$, which does not possess nontrivial real
solutions. Integrating the linear ODE arising in the brackets,
one obtains
\be\label{B-6}\varphi(x)=C_1e^x+C_2e^{-x}+\frac{d_2}{d_1}\,\psi.\ee
Substituting the  function $\varphi(x)$ from (\ref{B-6}) into the
second equation of system (\ref{B-4}), we arrive at the nonlinear
second-order ODE
 \be\label{B-7}\lf(C_1e^x+C_2e^{-x}+\frac{2d_2}{d_1}\psi\rg)\psi''+\frac{2d_2}{d_1}{\psi'}^2+
2\lf(C_1e^x-C_2e^{-x}\rg)\psi'-\frac{d_2}{d_1}\psi^2=0.\ee It turns
out that   this ODE can be integrated
by using  symbolic software packages (we used  Maple). One can also
reduce Eq.\,(\ref{B-7}) to a first-order ODE, which is solvable in a
straightforward way.  Thus, solving ODE (\ref{B-7}) and taking into
account (\ref{B-6}), we obtain the four-parameter family of
solutions \be\label{B-8}\ba
\varphi(x)=\frac{1}{2}\,\lf(C_1e^x+C_2e^{-x}\rg)\pm\frac{1}{2}\sqrt{\lf(C_1e^x+C_2e^{-x}\rg)^2+C_3e^x+C_4e^{-x}},
\medskip \\
\psi(x)=-\frac{d_1}{2d_2}\,\lf(C_1e^x+C_2e^{-x}\rg)\pm\frac{d_1}{2d_2}\sqrt{\lf(C_1e^x+C_2e^{-x}\rg)^2+C_3e^x+C_4e^{-x}}
\ea\ee of the ODE system (\ref{B-4}).

Substituting (\ref{B-8}) into ansatz  (\ref{B-3}), renaming $C_1
\rightarrow 2a_3d_2KC_1,$ $C_2 \rightarrow 2a_3d_2KC_2,$ $C_3
\rightarrow 4a_3^2d_2^2K^2C_3,$ $C_4 \rightarrow 4a_3^2d_2^2K^2C_4$,
and taking into account the scale transformation (\ref{B-10}), we
finally obtain the formulae
\begin{small}
\be\label{B-11}\ba
u(t,x)=d_2\lf(C_1e^{\sqrt{\frac{a_3}{d_3}}\,x}+C_2e^{-\sqrt{\frac{a_3}{d_3}}\,x}
\pm\sqrt{\lf(C_1e^{\sqrt{\frac{a_3}{d_3}}\,x}+C_2e^{-\sqrt{\frac{a_3}{d_3}}\,x}\rg)^2+
C_3e^{\sqrt{\frac{a_3}{d_3}}\,x}+C_4e^{-\sqrt{\frac{a_3}{d_3}}\,x}}\rg)\times\medskip\\
\hskip3cm\exp\lf(\frac{a_3d_1d_2}{d_3(d_1+d_2)}\,t\rg)-
\frac{d_2^2}{d_3(d_1+d_2)K},\medskip\\
v(t,x)=d_1\lf(-C_1e^{\sqrt{\frac{a_3}{d_3}}\,x}-C_2e^{-\sqrt{\frac{a_3}{d_3}}\,x}
\pm\sqrt{\lf(C_1e^{\sqrt{\frac{a_3}{d_3}}\,x}+C_2e^{-\sqrt{\frac{a_3}{d_3}}\,x}\rg)^2+
C_3e^{\sqrt{\frac{a_3}{d_3}}\,x}+C_4e^{-\sqrt{\frac{a_3}{d_3}}\,x}}\rg)\times\medskip\\
\hskip3cm\exp\lf(\frac{a_3d_1d_2}{d_3(d_1+d_2)}\,t\rg)-\frac{d_1^2}{d_3(d_1+d_2)K},\ea\ee
\end{small}where $C_i \ (i=1,\dots,4)$ are arbitrary constants.
Thus, (\ref{B-11}) represents a four-parameter family of the exact
solutions  of system (\ref{B-2}) with
$a_1=\frac{a_3d_1^2}{d_3(d_1+d_2)}$ and
$a_2=\frac{a_3d_2^2}{d_3(d_1+d_2)}$. Because $u$ and $v$ are
concentrations of monomers, one should take appropriate signs of
$C_i \ (i=1,\dots,4)$ in order to guarantee  nonnegativity of the
concentrations.
 It can be achieved, for example, if one takes upper signs and
 sufficiently large  $C_3>0$ and $C_4>0$.

\section{Conclusion} \label{sec-5}
In this work, a simplification of the well-known
Shigesada--Kawasaki--Teramoto  model \cite{sh-ka-te} is under study
by means of the conditional symmetry method. A complete  set of
$Q$-conditional (nonclassical) symmetries of the simplified SKT
system (\ref{2-1}) is derived and presented in Table \ref{tab1}. All
the symmetries obtained are new and do not coincide with the Lie
symmetries derived in \cite{ch-dav-muzy-17}. It should be also
stressed that several $Q$-conditional symmetries of  (\ref{2-1})
possess  highly nontrivial structures, in particular they are
nonlinear with respect to the functions $u$ and/or $v$ (see  the
operators $Q^7_0, \ Q^7, \ Q^{10}, \  Q^{11}, \  Q_0^{11}$). Such
type conditional symmetries do not arise in the  case of the
diffusive Lotka--Volterra  system \cite{ch-dav-2022}. We have no
doubt that  nonlinearities in these $Q$-conditional symmetries arise
from  cross-diffusion terms in the SKT system (\ref{2-1}).

The most promising symmetries (from applicability point of view)
were used
 for construction of exact solutions of the system in question.
 As a result, several exact solutions were constructed. These solutions, in particular  (\ref{3-10}), (\ref{3-19*}), and (\ref{3-20}), are essentially new because they have other structure
 distinct from  those derived earlier in \cite{ch-dav-muzy-17, ch-myr-08, li-2022}. Possible biological interpretation of the  exact solutions with correctly-specified parameters  is presented. In particular, it is shown how cross-diffusion pressure determines the  asymptotic behaviour of the species $u$ and $v$.
 The
solutions obtained can also be used as test problems for estimating
the accuracy of approximate
 and numerical methods for solving relevant boundary value problems.

An alternative application of the system in question was
demonstrated. We  have shown that the reaction-diffusion  system
(\ref{2-1}) with  correctly-specified parameters is   related to the
polymerisation process. Moreover, the relevant system admits a
$Q$-conditional (nonclassical) symmetry as well, so that a  family
of highly nontrivial  solutions was constructed (see formulae
(\ref{B-11})).

Finally, we point out that $Q$-conditional (nonclassical) symmetry
of the SKT system in the general form (\ref{2-1*}) is still unknown
even in the case of the constant potential $W(x)$. In contrast to
the Lie symmetry analysis, the relevant system of determining
equations is much more complicated and involves more
nonlinearities.  It is a great  challenge to solve that system
without essential restrictions.

\section{Acknowledgement}

R.Ch. acknowledges that this research was funded by the British
Academy's Researchers at Risk Fellowships Programme. J.R.K.
gratefully acknowledges a Leverhulme Trust fellowship.

\appendix

\section{Appendix: Sketch of proof of Theorem~\ref{th3}}\label{A}

In order  to prove Theorem~\ref{th3},   one needs to solve
completely the system of determining
Eqs.\,(\ref{2-36})--(\ref{2-44}) under the restrictions (\ref{2-2})
and (\ref{2-2*}). System (\ref{2-36})--(\ref{2-44}) is an
overdetermined system of PDEs. At the first sight, this system is
nonintegrable because that involves several nonlinear PDEs with the
unknown functions $\xi, \ \eta^1$ and $\eta^2$. We remind the reader
that the classical algorithm for finding Lie symmetries leads to the
linear  system of determining equations. It is shown below that the
nonlinear system (\ref{2-36})--(\ref{2-44}) is completely integrable
and has 12 inequivalent solutions depending on  values of the system
coefficients.

 Let us start from
Eqs.\,(\ref{2-36}). One notes  that solutions of these equations
depend essentially on the parameters $d_1, \ d_2,  \ d_{12}$ and
$d_{21}$. So, three cases should  be analysed:

\hskip3cm $\emph{\textbf{Case a.}}$ $d_1d_2d_{12}d_{21}\neq0.$

\hskip3cm $\emph{\textbf{Case b.}}$ $d_{12}=0, \ d_2\neq0.$

\hskip3cm $\emph{\textbf{Case c.}}$ $d_2=0.$

Formally speaking,  two more  cases $d_{21}=0, \ d_1\neq0$ and
$d_1=0$ arises; however, they are equivalent to $\emph{\textbf{Cases
b}}$ and $\emph{\textbf{c}}$ respectively, since  the discrete
transformation
 \be\label{a-0} u\rightarrow v, \quad  v\rightarrow u \ee
  preserves
the form of (\ref{2-1}).

 Because  (\ref{2-36}) consist of linear noncoupled  ODEs one easily solves them and obtains:
 \be\label{a-1} \ba
\emph{\textbf{Case a.}} \
 \ba
  \eta^1=\frac{P^1(t,x,u)}{d_1d_2+d_1d_{21}u+d_2d_{12}v}+P^0(t,x,u),\medskip\\
 \eta^2=\frac{Q^1(t,x,v)}{d_1d_2+d_1d_{21}u+d_2d_{12}v}+Q^0(t,x,v);
 \ea \medskip \medskip \\
\emph{\textbf{Case b.}} \
 \ba
  \eta^1=P^1(t,x,u)v+P^0(t,x,u),\medskip\\
 \eta^2=\frac{Q^1(t,x,v)}{d_2+d_{21}u}+Q^0(t,x,v);
 \ea  \medskip \medskip \\
 \emph{\textbf{Case c.}} \
 \ba
 \eta^1=P^1(t,x,u)v+P^0(t,x,u),\medskip\\
 \eta^2=\frac{Q^1(t,x,v)}{u}+Q^0(t,x,v),
 \ea \ea
  \ee
where $P^0, \ P^1, \ Q^0$ and $Q^1$ are to-be-determined smooth
functions.

Substituting the functions $\eta^1$ and $\eta^2$ from (\ref{a-1})
into Eqs.\,(\ref{2-37})--(\ref{2-38}), one can identify the
structure of the functions $P^0, \ P^1$ and $Q^0, \ Q^1$ of  the
variables $u$ and $v$, respectively. This allows us to reduce the
initial system with multidimensional PDEs to one involving ODEs and
two-dimensional PDEs, which can be solved in a straightforward way.

 Here we analyse in detail  only
$\emph{\textbf{Case c,}}$ which is most complicated and leads to the
most of cases presented in Table~\ref{tab1}. Notably, the
restriction $d_1d_{21}\neq 0$ automatically  takes place in this
case (see (\ref{2-2})).

$\emph{Analysis of \textbf{Case c.}}$  The general solution of
Eqs.\,(\ref{2-36})--(\ref{2-38}) has the form
 \be\label{a-2} \ba
\eta^1=f^0(t,x)+f^1(t,x)u+f^2(t,x)uv,\medskip\\
\eta^2=g^0(t,x)+g^1(t,x)v-f^2(t,x)v^2+\frac{g^2(t,x)-f^0(t,x)\,v}{u},\ea\ee
where $f^j$ and $g^j \, (j=0,1,2)$  are to-be-determined smooth
functions. Substituting (\ref{a-2}) into
Eqs.\,(\ref{2-39})--(\ref{2-42}) and splitting the equations
obtained with respect to the exponents of $u$ and $v$,  one  arrives
at the system
\begin{eqnarray} \label{a-3} && f^0\xi=0, \ g^2\xi=0, \ d_{12}f^2\xi=0,\\
\label{a-4*} &&
 d_1f^2\xi-d_{12}\left(\xi_t+2\xi\xi_x+(g^1-f^1)\xi\right)=0,\\
\label{a-4} &&
\xi_t+2\xi\xi_x+\left(\frac{d_{12}}{d_1}\,g^0-f^1\right)\xi=0, \
\xi_t+2\xi\xi_x-2d_1g^1_x=0, \ 2d_1g^0_x-g^0\xi=0, \\ \label{a-5} &&
2d_1f^2_x+f^2\xi=0,  \ \xi_{xx}-2f^1_x-2g^1_x=0.
\end{eqnarray}
The latter is an overdetermined system for finding six functions,
which consists of three algebraic equations and six differential
equations. Because all algebraic  and two differential equations
vanish by setting $\xi=0$,
 we need to examine two essentially different possibilities,
$\xi\neq0$ and  $\xi=0$.

\textbf{\emph{Let us assume  $\xi\neq0$}}, then  the functions
$f^0=f^2=g^2=0$ are immediately found from  Eqs.\,(\ref{a-3}).
Obviously, Eq.\,(\ref{a-4*}) leads to two different subcases
 $d_{12}\neq0$ and $d_{12}=0$.

\emph{Analysis of subcase $\xi\neq0, \ d_{12}\neq0.$} Analysing
Eqs.\,(\ref{a-4*}) and (\ref{a-4}), the relations
 $g^1=\frac{d_{12}}{d_1}\,g^0$ and
$f^1=\frac{2d_{12}}{d_1}\,g^0$ between the functions $g^0, \ g^1$
and $f^1$ were identified.  Now Eqs.\,(\ref{a-4})--(\ref{a-5}) can
be rewritten in the form \be \label{a-8}
 \xi_t+2\xi\xi_x-\frac{d_{12}}{d_1}g^0\xi=0, \ 2d_1g^0_x-g^0\xi=0, \
\xi_{xx}-6\frac{d_{12}}{d_1}g^0_x=0. \ee

It turns out  that any solution of the overdetermined system
(\ref{a-8}) leads to a $Q$-conditional symmetry operator, which is
equivalent to the Lie symmetry operator of the SKT system
(\ref{2-1}).
 Let us show this. Integrating the third equation of system
(\ref{a-8}), one obtains $g^0=\frac{d_1}{6d_{12}}\,\xi_x+h(t)$
($h(t)$ is an arbitrary function). So, the first two equations of
(\ref{a-8}) are reducible to
\be\label{a-29}\xi_t=-\frac{11}{6}\xi\xi_x+\frac{d_{12}}{d_1}\,h\xi,
\quad
\xi_{xx}=\frac{1}{2d_1}\xi\xi_x+\frac{3d_{12}}{d_1^2}\,h\xi.\ee
 Now we analyse the identity
$\xi_{txx}=\xi_{xxt}$ using Eqs.\,(\ref{a-29}) and their
differential consequences.
 As a result,   the equation
\[\xi_{xx}\left(11d_1\xi_{x}+51d_{12}h\right)=0\]
was derived that immediately gives  $\xi_{xx}=0$. The latter leads
to   $g^0=0$ (see the last two equations in (\ref{a-8})). Thus, the
$Q$-conditional symmetry operator (\ref{2-35}) takes the form
\[ Q=\p_t+\xi(t,x)\,\p_x,\]
 where the function $\xi$ is any  solution
of the system \[\xi_t+2\xi\xi_x=0, \ \xi_{xx}=0,\]
solutions of  which are readily defined: the trivial solution
$\xi=const$ leads  to Lie symmetry operator $\p_t+\alpha\p_x$;  the
only  nontrivial solution is $\xi=\frac{x+\alpha_1}{2t+\alpha_0}$
($\alpha_1$  and $\alpha_0$ are arbitrary constants), so that
 the
operator $Q^*=\p_t+\frac{x+\alpha_1}{2t+\alpha_0}\p_x$  results.
 To complete
the examination of \emph{subcase $\xi\neq0, \ d_{12}\neq0$}, we need
to examine the last two equations of the system  of determining
equations.
 Substituting the functions $\xi=\frac{x+\alpha_1}{2t+\alpha_0}, \
\eta^1=0$ and $\eta^2=0$ into Eqs.\,(\ref{2-43})--(\ref{2-44}), we
arrive at the conditions: $a_1=a_2=b_1=b_2=c_1=c_2=0$. This means
that the SKT system (\ref{2-1}) simplifies to the form
\[\ba u_t = [(d_1+d_{12}v)u]_{xx},\\
 v_t = d_{21}[uv]_{xx}.\ea\]
 The latter admits the   Lie symmetry $(2t+\alpha_0)\p_t+(x+\alpha_1)\p_x$
 (see p.\,88 in \cite{ch-dav-muzy-17}). On the other hand, the operator $Q^*$ is equivalent
  to the above Lie symmetry because each conditional symmetry is defined up to an
  arbitrary  multiplier $M$, in paticular $M=2t+\alpha_0$.

\emph{Subcase} $\xi\neq0, \ d_{12}=0$ can be studied in a quite
similar way and that also leads only to the Lie symmetry operators
listed above.

\textbf{\emph{In the case $\xi=0$}}, Eqs.\,(\ref{a-3})--(\ref{a-5})
reduce  to the trivial equations   $f^1_x=f^2_x=g^0_x=g^1_x=0.$
Thus, formulae (\ref{a-2}) take the form
 \be\label{a-9}
\eta^1=f^0(t,x)+f^1(t)u+f^2(t)uv,\
\eta^2=g^0(t)+g^1(t)v-f^2(t)v^2+\frac{g^2(t,x)-f^0(t,x)\,v}{u}.\ee
Substituting (\ref{a-9}) into Eqs.\,(\ref{2-43})--(\ref{2-44}) and
splitting the equations obtained with respect to the exponents  of
$u$ and $v$, one immediately finds  $f^0=0$. The other five
functions
follow  by solving  the system
\begin{eqnarray} \label{a-10} && b_1g^0=0, \ b_1f^2=0, \
b_1(d_1f^1-d_{12}g^0)=0, \\ \label{a-13} && (2c_2-3f^2)g^2=0,  \
(c_2-f^2)f^2=0, \ (d_1^{\,2}f^2-d_{12}^{\,2}g^0+d_1d_{12}g^1)g^2=0,
\\ \label{a-11} &&  (b_2d_1+a_1d_{21}-d_{21}f^1)g^0=0, \
(b_2d_{12}-c_1d_{21}-d_{21}f^2)g^0=0, \\  \label{a-12} &&
(c_2-f^2)(d_1^{\,2}f^2-d_{12}^{\,2}g^0+d_1d_{12}g^1)=0, \
(a_2-g^1)(d_1^{\,2}f^2-d_{12}^{\,2}g^0+d_1d_{12}g^1)=0, \\
 \label{a-14} &&
\frac{df^1}{dt}+\left(c_1+f^2+\frac{d_{12}}{d_1}\,(a_1-f^1)\right)g^0=0, \\
\label{a-15} &&
d_{21}\frac{df^2}{dt}-(b_2d_1+a_1d_{21}-d_{21}f^1)f^2-(b_2d_{12}-c_1d_{21}-d_{21}f^2)g^1=0,
\\ \label{a-16} &&
\frac{df^2}{dt}+(c_2-f^2)\left(f^1-g^1-\frac{d_{12}}{d_1}\,g^0\right)-(a_2-g^1)f^2=0,
\\ \label{a-18} &&
d_{12}\frac{dg^0}{dt}+(d_1f^2-d_{12}(a_2-g^1+f^1))g^0+(b_2d_{12}-c_1d_{21}
-d_{21}f^2)g^2=0,
\\ \label{a-17} &&
\frac{dg^1}{dt}+(a_2-g^1)f^1+\left(2(c_2-f^2)-\frac{d_{12}}{d_1}\,(a_2-g^1)\right)g^0=0,
\\  \label{a-19*} && g^2_t-\left(a_2+2f^1-\frac{d_{12}}{d_1}\,g^0\right)g^2=0,
\\  \label{a-19} &&
d_{21}g^2_{xx}-b_2g^2-\frac{dg^0}{dt}+\left(a_2+f^1-\frac{d_{12}}{d_1}\,g^0\right)g^0=0.
\end{eqnarray}
Eqs.\,(\ref{a-10})--(\ref{a-12}) are algebraic, while
Eqs.\,(\ref{a-14})--(\ref{a-19}) form a seven-component system of
differential equations. A simple analysis of Eqs.\,(\ref{a-10})
highlights   two subcases  $b_1\neq0$ and $b_1=0$, which lead to
essentially different results.

 \emph{In the subcase} $b_1\neq0$ one immediately finds the functions  $g^0=f^1=f^2=0$, so that  formulae (\ref{a-9}) take the form
\be\label{a-20} \eta^1=0, \ \eta^2=g^1(t)v+\frac{g^2(t,x)}{u}.\ee
Simultaneously  Eqs.\,(\ref{a-13})--(\ref{a-19}) are  simplified to
the algebraic  equations
\begin{eqnarray} \label{a-21} && c_2g^1=0, \ c_2g^2=0, \
(b_2d_{12}-c_1d_{21})g^1=0, \ (b_2d_{12}-c_1d_{21})g^2=0, \\
\label{a-22} && d_{12}(a_2-g^1)g^1=0, \ d_{12}g^1g^2=0,
\end{eqnarray}
and two linear PDEs \be\label{a-23} g^2_t-a_2g^2=0, \
d_{21}g^2_{xx}-b_2g^2=0. \ee

As one can note from Eqs.\,(\ref{a-21}), the restrictions $c_2=0$
and $b_2d_{12}-c_1d_{21}=0$  should hold (otherwise the Lie symmetry
operator $Q=\p_t$ is obtained). Note that the parameter $d_{12}$
must be nonzero, otherwise $c_1=0$ and  a contradiction to
(\ref{2-2*}) is obtained. The remaining
Eqs.\,(\ref{a-22})--(\ref{a-23}) have the solutions
 \[ g^1=a_2, \quad g^2=0 \]
 and
  \[ g^1=0, \quad g^2= e^{a_2t}g(x), \  d_{21}g_{xx}-b_2g=0. \]
  Thus, inserting the above functions into formulae (\ref{a-20}), one obtains two $Q$-conditional symmetry operators for the
  SKT system
\be\label{a-24}\ba u_t = [(d_1+d_{12}v)u]_{xx}+u\left(a_1-b_1u-\frac{d_{12}}{d_{21}}\,b_2v\right),\\
 v_t = d_{21}[uv]_{xx}+v(a_2-b_2u),\ea\ee where $d_1, \ d_{12}$ and
 $d_{21}$ are nonzero constants.
 It can be noted that the scaling  transformation
\be\label{a-25} t \rightarrow \frac{1}{d_1}\,t, \
u\rightarrow\frac{d_1}{d_{21}}\,u, \
v\rightarrow\frac{d_1}{d_{12}}\,v\ee essentially simplifies system
(\ref{a-24}) and the operators obtained.  Finally,  the system and
the conditional symmetries $Q^2$ and  $Q_0^3$
 presented in Case 2 of Table~\ref{tab1} are derived.


\emph{ In the   subcase} $b_1=0$, a much larger set of   conditional
symmetry operators was  obtained because  Eqs.\,(\ref{a-10}) vanish.

 Analysis of Eqs.\,(\ref{a-13}) under \emph{the assumption}  $g^2\neq0$
 leads to the conditions $c_2=0$ and $f^2=0$ (see first two equations in (\ref{a-13})).
 Using the second equation
of (\ref{a-11}) and Eq.\,(\ref{a-18}), we obtain
$b_2d_{12}-c_1d_{21}=0$, hence again $d_{12}\neq0.$ So,  applying
transformation (\ref{a-25}), the nonzero parameters are reducible to
$d_1=d_{12}=d_{21}=1$. As a result,
Eqs.\,(\ref{a-13})--(\ref{a-19}) can be simplified to the form
\begin{eqnarray} \nonumber && g^1=g^0, \  \frac{df^1}{dt}=0, \ a_2f^1=0, \ (a_1+b_2-f^1)g^0=0,\\
\nonumber && \frac{dg^0}{dt}=(a_2+f^1-g^0)g^0, \\
\nonumber && g^2_t=(a_2+2f^1-g^0)g^2, \ g^2_{xx}-b_2g^2=0.
\end{eqnarray}

The general solution of the above system depends on  relations
between the parameters $a_1, \ a_2$ and $b_2$. All possible
relations were examined. So, five new operators and the relevant SKT
systems  arising in   Table~\ref{tab1} were identified,
namely\,: \\
 $1)$ $a_2\neq0$ and $a_1+b_2\neq0$: operator $Q_0^3$ (see Case 7 of
 Table~\ref{tab1}); \\
  $2)$ $a_2\neq0$ and $a_1+b_2=0$: operator $Q^3$ (see Case 10 of
 Table~\ref{tab1}); \\
$3)$ $a_2=0$ and $a_1+b_2\neq0$: operators $Q_0^{11}$ (if $f^1\neq
a_1+b_2$) and $Q^{11}$ (if $f^1=a_1+b_2$) (see Case 9 of
 Table~\ref{tab1}); \\
$4)$ $a_2=0$ and $a_1+b_2=0$: operators $Q_0^{11}$ (if $f^1\neq 0$)
and $Q^{12}$ (if $f^1=0$) (see Case 11 of
 Table~\ref{tab1}).

  Under \emph{the assumption}  $g^2=0$,  Eqs.\,(\ref{a-13})--(\ref{a-19})
   were   analysed as well and the  operators
$Q^1, \ Q^4,$ $Q_0^7,$ $Q^7, \ Q^8, \ Q^9$ and $Q^{10}$  and the
corresponding  systems arising in Table~\ref{tab1} were found.

 Thus, $\emph{\textbf{Case\,c}}$ was  completely examined and
Cases 1, \ 2, \ 3, \ 6--11 of Table~\ref{tab1} were obtained.

$\emph{\textbf{Cases\,a}}$ and  $\emph{\textbf{b}}$
 have been investigated in quite a  similar way. As a result, Cases 4, \ 5 and 12 of Table~\ref{tab1} were identified.
Note that $\emph{\textbf{Case\,b}}$ leads
 only to the SKT systems with $d_{12}=c_1=0$, which are excluded
from  consideration (see the first restriction in (\ref{2-2*})).

The set of local transformations (\ref{2-38}) was obtained by
incorporation of transformations of the form  (\ref{a-0}) and
(\ref{a-25}), which were used in order to simplify and unify the
results.

The sketch of the proof is now complete.


\end{document}